\documentclass{jfm}
\usepackage{graphicx}
\usepackage{epstopdf, epsfig}

\usepackage{xcolor}

\usepackage{caption}
\usepackage{subcaption}
\usepackage{amsmath}
\usepackage{amssymb}
\usepackage{sidecap}

\shorttitle{Turbulent Particle Pair Diffusion}
\shortauthor{N. A. Malik \& F. Hussain}

\title{Objective quantification of  Particle Pair Diffusion in Homogeneous Isotropic Turbulence}
         
\author{Nadeem A. Malik\corresp{\email{nadeem.malik@ttu.edu and nadeem\_malik@cantab.net}} and Fazle Hussain}

\affiliation{Department of Mechanical Engineering, Texas Tech University, Lubbock, Texas 74909, USA}

\begin{document}

\maketitle

\begin{abstract}
Turbulence consists of interacting flow structures covering a wide range of length and time scales. A long-standing question looms over pair diffusion of particles in close proximity i.e. particle pair diffusion at small separations: what range of turbulence length scales governs pair diffusion? Here, we attempt to answer this question by addressing pair diffusion by both fine scales and larger scale coherent structures in which the fine scales are embedded - we unavoidably encounter a combination of both local and non-local interactions associated with the small and large length scales. The local structures possess length scales of the same order of magnitude as the pair separation $l$, and they induce strong relative motion between the particle pair. However, the non-local structures, possessing length scales much larger than $l$, also induce (via Biot-Savart) significant relative motion, an effect ignored in prior studies (based on Richardson-Obukhov R-O theory). This fundamentally changes the interpretation of the pair diffusion process, giving the pair diffusivity $K$ scaling as $K\sim l^{1.556}$ -- agreeing within $1\%$ of data. The `R-O constant' $g_l$ is shown to be not a constant, although widely assumed to be a constant. However, new constants (representing pair diffusivity $G_K$  and pair separation $G_l$) are identified, which we show to asymptote to  $G_K\approx 0.73$ and $G_l\sim 0.01$ at high Reynolds numbers. As an application, we show that the radius of a cloud of droplets in a spray is smaller by an order of magnitude as compared to R-O theory.

\end{abstract}


\section{Introduction}\label{Sec1}

\subsection{Turbulent pair diffusion}\label{SS1.1}
Transport  is  prevalent in natural and industrial flows. Analysis of fluid particle separation is a route to understanding these phenomena and was introduced by \cite{Richardson1926}. Numerous research (e.g. \cite{Durbin1980,Thalabard2014,Darragh2020,Anisimov2020}) have studied pair diffusion, almost all of them based on a locality hypothesis (explained below), but many issues remain unresolved.

\begin{figure}
   \centerline{\includegraphics[width=12cm]{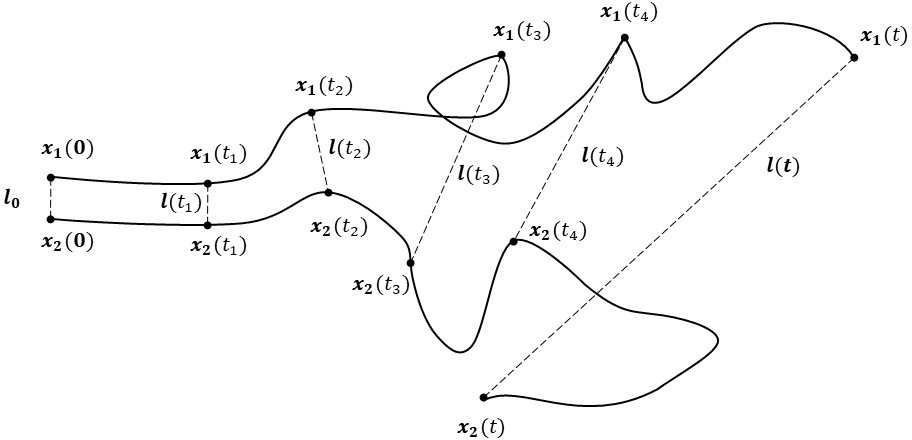}}
   \caption{A sketch of pair separation in time}    \label{fig_01}
\end{figure}

There are several different regimes of pair diffusion, associated with the viscous subrange, the inertial subrange, and the large scales. Here we address pair diffusion in the inertial subrange. Fig. \ref{fig_01}  illustrates the time development of a pair of particles released in a field of homogeneous and isotropic turbulence with a small separation $l_0$ (of the order of the Kolmogorov scale $\eta$). Richardson surmised from particle motion in atmospheric turbulence that the average rate of pair separation increases with increasing separation because stronger winds with more energy occur at bigger scales. From data of turbulent pair diffusion coefficients (diffusivities)  collected from different geophysical and laboratory observations, Richardson assumed a scale dependent pair diffusion coefficient $K\left(l\right)$ across all scales $l$ of separation and intuited a constant power law fit to the data, 
\begin{eqnarray}\label{eq_1.0}
   K\left(l\right) &\sim& l^{4/3}, 
\end{eqnarray}
where $l(t)=|{\bf x}_2(t)-{\bf x}_1(t)|$ is the distance between the two particles located at ${\bf x}_1$ and ${\bf x}_2$ at time $t$, and $K\left(l\right)$ is the ensemble averaged quantity $K\left(l\right)=\langle {\bf l}(t)\cdot {\bf v}(t)\rangle$, ${\bf l}(t)$ is the  pair separation vector and ${\bf v}({t})={\bf u}_2(t)-{\bf u}_1(t)$ is the pair relative velocity; ${\bf u}({\bf x},t)$ is the turbulent velocity field, and the particle velocities at time t are, respectively, ${\bf u}_1 (t)={\bf u}({\bf x}_1 (t),t)$ and ${\bf u}_2 (t)={\bf u}({\bf x}_2 (t),t)$.
Richardson's $4/3-$law is equivalent to, \cite{Obukhov1941}, 
\begin{eqnarray}\label{eq_1.1}
    \langle l^2 \rangle &=& l_0^2 + g_l\varepsilon t^3,
\end{eqnarray}
in the inertial subrange often referred to as the Richardson-Obukhov (R-O) $t^3$-regime, $\varepsilon$ is the rate of kinetic energy dissipation per unit mass. $g_l$ is presumed to be a universal constant. The R-O regime is assumed valid in infinite Reynolds number turbulence containing an infinite inertial subrange.

Richardson assumed that pair diffusion can be modelled by a diffusion equation with the scale dependent diffusion coefficient. There is no proof that such a formulation is consistent with the Navier-Stokes equation except for an ensemble of particle pairs released with the same initial separation $l_0$ (point source release) within the inertial subrange, $\eta\le l_0\ll L_1$, where the Kolmogorov scale $\eta$ and $L_1$ are the lower and upper scales of the inertial subrange. Richardson formulated his theory in terms of a joint probability density function for the pair separation, $q(r,t)$, which satisfies the 3D diffusion equation (\cite{Richardson1926, Salazar2009}),
\begin{eqnarray}\label{eq_1.2}
   \frac{\partial q}{\partial t} &=& \frac{1}{r^2} \frac{\partial }{\partial r}\left({ r^2 K(r) \frac{\partial q}{\partial r}}\right).
\end{eqnarray}
where $r$ is the pair separation variable,  subject to the normalization, $\int_0^{\infty} 4\pi r^2 q(r,t) dr=1$.

{The diffusivity defines the effective rate of spread of a `substance’ in a fluid medium; for example, turbulent diffusivity is orders of magnitude greater than molecular diffusivity. Note that although the R-O constant $g_l$ is assumed to be a constant,  there is as yet no consensus on its value.

Richardson's assumed 4/3-scaling law is based on a concept of locality which is consistent with Kolmogorov  K41 theory. By locality is meant that the increase in the pair separation distance $l$ depends on turbulence motions of similar length scale  $l$. Thus,
the energy spectrum in the inertial subrange is $E(l)\sim \varepsilon^{2/3} l^{5/3}$,  and the typical pair relative velocity at separation $l$ is $v(l)\sim \sqrt{E(l)/l}$, and from the scaling for the turbulent diffusion coefficient $K(l)\sim lv(l) \sim \sqrt{lE(l)}$, it trivially follows that $K(l) \sim \varepsilon^{1/3} l^{4/3}$ -- this $l^{4/3}$ dependence is the key result of  R-O's locality theory, valid in the limit of very high Reynolds numbers. 

The existence of a solution from a point source implies that the initial separation of particle pairs is effectively zero, and therefore the Kolmogorov scale must also asymptote to zero. Richardson's theory is thus strictly applicable only in the asymptotic limit of infinite Reynolds number, $Re\to\infty$ (hence infinite inertial subrange), and also that $l_0\to 0$, as $\eta\to 0$, while keeping $l_0/\eta \ge 1$.

It is possible to generalize the scaling for the diffusion coefficient to be time dependent and still be consistent with K41 (\cite{Klafter1987,Salazar2009}) such that	$K \sim \varepsilon^a t^b l^c$ for real $a,b$,and $c$.  However, a time dependent diffusion coefficient is hard to justify physically if we assume steady state equilibrium turbulence, and we do not consider it further here.

Until recently, almost all theories of pair diffusion have assumed R-O locality and the existence of the R-O $t^3$ regime (see equation (\ref{eq_1.1})). And yet, surprisingly, the collection of  experimental data and Direct Numerical Simulations, is still not convincing because of the relatively low Reynolds numbers in the experiments and DNS (contrary to the assumption in the R-O theory), and the high error levels in the data in many experiments. This has led \cite{Salazar2009} to note that, " .. there has not been an experiment that has unequivocally confirmed R-O scaling over a broad-enough range of time and with sufficient accuracy." 

Furthermore, a re-examination of  Richardson's original 1926 collection of datasets reveals that one of the datapoints came from molecular diffusion experiments which are not relevant to turbulence studies; removing this datapoint reveals that Richardson's $K\sim l^{4/3}$ local scaling is a rather poor fit, but $K\sim l^{1.564}$ scaling is an almost exact fit to the data -- this difference in the power from $4/3$ to $1.564$ is rather significant. This clearly points to the failure of the assumption of locality in the R-O theory. 

Hereafter, we follow the usual convention of replacing scaling on $l$ with scaling with its rms value, that is $l^2=\langle l^2(t)\rangle$.
}

The rest of the paper is organized as follows. Section 2 discusses the physical meaning of local and non-local neighbourhoods of flow structures. Section 3  summaries the newly developed non-local theory and its predictions for the pair diffusion coefficient scaling laws.
Section 4 reviews the current state of knowledge on the R-O constant, and we show that in the non-local theory it is not a constant which explains the wide scatter in its estimate. However, new constants emerge and their asymptotic behavior in the limits of small and large inertial subranges $R_k$ is discussed. The quantitative effect that the non-local theory could have on particle spread in real world problems (e.g. dispersal in a spray) is estimated. In Section 5, we  investigate the effect that the frequency (the rate at which flow structures change) have on pair diffusion. In the discussions in Section 6,  we argue that these results fundamentally change our interpretation of the pair diffusion process which impacts on turbulent diffusion modelling and prediction in real world applications.

%%%%%%%%%%%%%%%%

\section{\textbf Local and non-local neighbourhoods of turbulent motions}\label{Sec2}

{The formal problem is to determine the pair diffusivity, $K=\langle {\bf l\cdot v}\rangle$, of an ensemble of pairs of fluid particles in a field of homogeneous turbulence containing an extended inertial subrange with an energy density spectrum, $E(k)$, such that $E(k)\to 0$ as $k\to\infty$. We assume point source release, which in practical terms means that the initial pair separation,  at some earlier time $t_0$, denoted by $l_0=|{\bf x}_2 (t_0 )- {\bf x}_1 (t_0)|$, is close to or greater than the Kolmogorov length scale  but inside the inertial subrange, $\eta \le l_0 \ll L_1$. The particles will diffuse apart and eventually decorrelate with the initial conditions -- they will 'forget' their initial conditions, $(l_0,v_0)$, as Batchelor put it -- after some travel time, $t_{l_0}$, inside the inertial subrange. During this travel time, the pair will display ballistic motion with essentially constant relative velocity equal to its initial value $v=v_0$; 
$ l^2=l_0^2+v_0^2(t-t_0)^2$. The transition from the ballistic regime to the explosive inertial subrange regime occurs on a time scale of the order of the eddy turnover time scale,  $t_{l_0}\approx \tau_0(l_0) \sim \varepsilon^{1/3} l_0^{1/3}$. If $l_0\approx \eta$ then the ballistic travel time is approximately equal to the Kolmogorov time microscale, $t_{l_0}\approx \varepsilon^{1/3} \eta^{1/3}$, which is very short.  At much longer times we can ignore the ballistic regime because $t\gg t_{l_0}$ and $l\gg l_0$. Below we focus on pair diffusion well inside the inertial subrange where these conditions prevail, and without loss of generality we let $t_0=0$, and ignore the ballistic regime.

We consider a field of homogeneous isotropic turbulence with generalized energy spectrum,
\begin{eqnarray}\label{eq_2.2}
    E(k)  &\sim& L^{5/3-p}\varepsilon^{2/3}\ k^{-p},  \quad  {\rm for\ } 1<p\le3
\end{eqnarray}
in the wavenumber range $k_1\le k\le k_\eta$; $k_1=1/L_1$ and $k_\eta=1/\eta$ are respectively the wavenumbers corresponding to the largest scale in the inertial subrange and the Kolmogorov scale. 
$L\ge L_1$ is some length scale characteristic of the very large scales of motions outside which contains most of the kinetic energy and is needed for dimensional consistency. $L$ is not necessarily equal to $L_1$, but a comparison between different systems would need an agreed convention on how $L$ scales with $L_1$.

A convenient measure of the size of the inertial subrange $R_k$ is,
\begin{eqnarray}\label{eq_2.3} 
    R_k &=& \frac{k_\eta}{k_1}.                                                                    
\end{eqnarray}
$R_k$ can serve in place of Reynolds number if  $L_1$ (hence $L$) could be related to a more usual length-scale such as the integral length scale $L_I$ or the Taylor microscale $\lambda_T$. To illustrate, let us suppose that $L_I/L_1\approx 10$, then from the relationship $Re\sim \left(L_I/\eta\right)^{4/3}\sim \left(10R_k\right)^{4/3}$, we obtain $R_k\sim {Re}^{3/4}/10$. We will show that using $L_1$ and $R_k$ greatly simplifies matters and they play important roles in the new picture.
}

\begin{figure}
\begin{center}
\begin{subfigure}{0.9\textwidth}
   \centerline{\includegraphics[width=7cm]{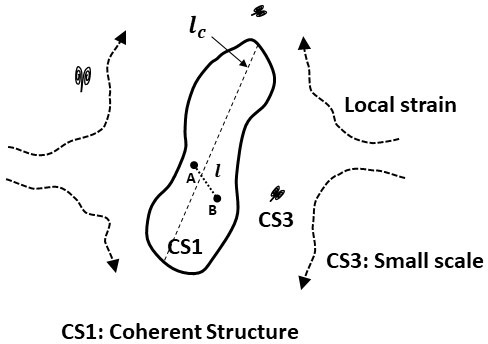}}
   \caption{Richardson-Obukhov locality: the relative motion is determined only by flow structure CS1 of comparable size to the pair separation $l_c\sim l$.}
\end{subfigure}\\
\vspace{3mm}
\begin{subfigure}{0.9\textwidth}
   \centerline{\includegraphics[width=10cm]{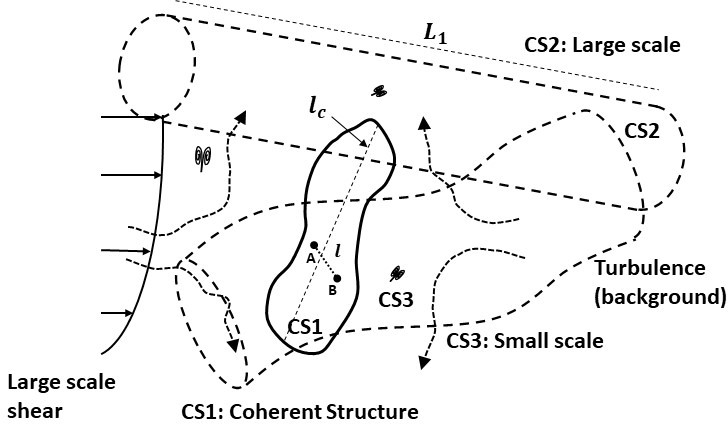}}
   \caption{Non-local theory: both CS1 and CS2 (size $L_1> l_c$) determine the relative motion.}
\end{subfigure}\\ 
\vspace{3mm}
\begin{subfigure}{0.9\textwidth}
   \centerline{\includegraphics[width=9cm]{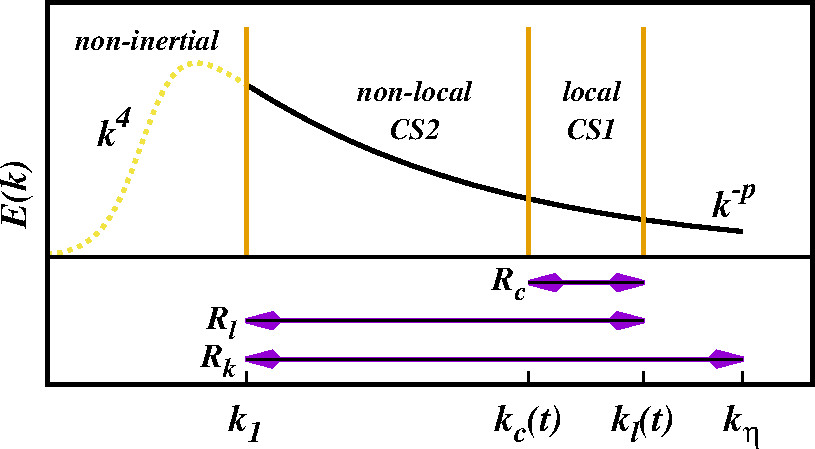}}
   \caption{Energy spectrum showing the local and non-local neighbourhoods to the pair separation wavenumber $k_l(t)\sim 1/l(t)$ separated by the vertical line at $k_c(t)$. $k_l(t)$ and $k_c(t)$ are time-dependent, but $R_c=k_l/k_c$ is assumed constant, equation (\ref{eq_2.1}). $k_1=1/L_1$,  $R_k=k_\eta/k_1$ and $R_l=k_l/k_1=L_1/l$.}
\end{subfigure}
   \caption{(Colour online). {Illustrations of a particle pair A-B (dots) in a field of turbulence and subjected to random shearing and straining.} }\label{fig_02}
\end{center}
\end{figure}

{Central to the new theory is the concept of local and non-local  neighbourhoods of turbulent motions with respect to the pair separation distance. We first rationalize these concepts through formal and general arguments based on flow structure. After this, we will summarize the non-local theory and its main predictions in Section 3.

In a body of homogeneous isotropic turbulence there exist some very large scale structures of the order of the integral length scale, or even larger, possessing length scales $L$ and velocity scale $U$.  
There also exist turbulent motions of comparable size to the pair separation $l$, and turbulent motions of intermediate  size. 

We first note what non-locality {\em does not} mean. The very large sized  motions induce (via Biot-Savart) relative velocities at the small separations $l\ll L$ which are much smaller than the local pair relative velocity $v(l)$, i.e. $l(dU/dx)\ll v(l)$, so they contribute negligible additional relative motion to the pair. Thus, the term non-local excludes the influence of  such very large size motions which simply carry or `sweep' the pair along in the flow. 

With respect to the pair separation $l$, the term non-local (or non-local neighbourhood) therefore pertains to turbulent motions in a range of sizes  $L_1\ll l\ll l_c$ (which corresponds to the wavenumber range $k_1\le k< k_c$ in Fig. \ref{fig_02}(c)).

Formally, the relative velocity ${\bf v}$ across a finite displacement ${\bf l}$ in the fine scales of turbulence is,
\begin{eqnarray}\label{eq_10}
	{\bf v}({\bf l}) &=& {\bf u}({\bf x}_2 )-{\bf u}({\bf x}_1).
\end{eqnarray}
Using the Fourier decomposition of the velocity field, ${\bf u}({\bf x}) = \int {\bf A(k)}    \exp\left({ i{\bf k\cdot x}}\right) \  d{\bf k}$,  where ${\bf A(k)}$ is the Fourier transform of the flow field, and ${\bf k}$ is the associated wavenumber, \cite{Batchelor1953}, a
Fourier decomposition of the relative velocity can also be derived, averaging over all space due to homogeneity, yields
\begin{eqnarray}\label{eq_11}
 {\bf v}({\bf l}) &=& \int {\bf A(k)}    [\exp\left({ i{\bf k\cdot l}}\right)-1] \  d{\bf k}.
\end{eqnarray}
Taking the scalar product of ${\bf v}$ with ${\bf l}$ and then its ensemble average $\langle\cdot\rangle$ over particle pairs yields the pair diffusivity,
\begin{eqnarray}\label{eq_12}
  K(l) &\sim& \langle{\bf l\cdot v}\rangle  
  \sim \int \langle{\bf (l\cdot A)} [\exp\left({ i{\bf k\cdot l}}\right)-1] \rangle d{\bf k}.
\end{eqnarray}
We partition this expression as the sum of three integrals,
\begin{align}\label{eq_15}
  K(l) &\sim \left({\int_{nl}+ \int_l+ \int_s}\right) \langle ({\bf l\cdot A})
       (\exp\left({ i{\bf k\cdot l}}\right)-1)\rangle \ d {\bf k}
\end{align}
which we rephrase as,
\begin{eqnarray}\label{eq_16}
	K(l) &\sim& K^{nl}+K^l+K^s.
\end{eqnarray}
The ranges of integration in each of the three integrals are contiguous, to be defined in Section 3. 

Physically, the contribution $K^s$ comes from the scales of motion smaller than the pair separation itself, so the integral for $K^s$ is taken over the wavenumber range $ k\ge k_l$.
The energy contained in these scales is very small if the energy spectrum decreases as $k$ increases, such as an inverse power law of the type $E(k)\sim k^{-p}$, with $p>1$. 
Furthermore, these small scale motions possess very small time scales and hence  very high frequencies which induce rapid changes in the direction and magnitude of the relative velocity vector, effectively randomizing their motions like a random walk type of process. 
The net statistical impact of the very small scales is therefore expected to be very small,  $K^s\ll Max(K^{nl},K^l)$, and henceforth we will ignore $K^s$. This leaves just the contributions $K^{nl}$ and $K^l$. 

For a physical understanding of  local and non-local processes and neighbourhoods it is convenient to consider the diffusional processes in both physical space and wavenumber space.  Fig. \ref{fig_02}(a) shows a schematic of the R-O locality interpretation of particle pair growth in a field of turbulence, where a particle pair A-B has separation $l$. In this picture, only flow structures CS1  whose length scale is of the same order of magnitude as $l$, i.e. $l_c\sim l$, induce an influence on the pair separation growth.

Fig. \ref{fig_02}(b) shows a schematic of the interpretation of pair separation growth in the non-local picture;
a field of turbulence contains flow structures of all sizes, and flow structures CS1 and CS2 (whose length scale $L_1$ are bigger than $l_c$, i.e. $L_1> l_c$) could influence the pair separation growth. (Coherent structures CS3 of small length scale also exist, but do not contribute to the pair growth.)
  
In the wavenumber space, this is equivalent statistically to Fig. \ref{fig_02}(c) which shows a typical  energy spectrum which includes the very large scales where experiments have determined that $E(k)\sim k^4$, and the small scales where $E(k)\sim k^{-p}$, with a smooth fit between them. Because we will be considering turbulence with intermittency corrections as well as Kolmogorov turbulence which have different power laws, we will consider general spectra with $1<p\le 3$ . The inertial subrange is the wavenumber range $k_1\le k\le k_\eta$, whose size is defined by equation (\ref{eq_2.3}). The pair separation wavenumber is 
\begin{eqnarray}\label{eq_2.8b}
    k_l (t) &\sim& \frac{1}{l(t)},
\end{eqnarray}
and since we focus on separations well inside the subrange, we have $\eta\ll l\ll L_1$, (which is equivalent to $k_1\ll k_l\ll k_\eta$). We define the size of the inertial subrange relative to the pair separation,
 \begin{eqnarray}\label{eq_2.8c}
    R_l (t) &=& \frac{k_l}{k_1},
\end{eqnarray}
which is an important quantity in later analysis. The wavenumber that separates the two groups of CS's is $k_c(t)$ in Fig. \ref{fig_02}(c), and
 \begin{eqnarray}\label{eq_2.8d}
    k_c (t) &\sim& \frac{1}{l_c(t)},
\end{eqnarray}
such that $k_1\ll k_c\ll k_l$. The range of wavenumbers  in the local neighbourhood is, $k_c\le k\le k_l$.

Consider a pair of particles A-B inside a flow structure CS1 and subject to turbulence and random straining and shear, depicted in Fig \ref{fig_02}(b). The pair relative motion is strongly correlated to the local flow structure CS1 ($l_c\sim l$) and we say that the pair is interacting locally with CS1. 
$l_c$ clearly defines the neighbourhood of locality, corresponding to the bandwidth wavenumber range $k_c<k\le k_l$ in Fig. \ref{fig_02}(c).

A turbulent flow also contains CS2, of much bigger length scale than $l_c$ and smaller than $L_1$, which induce motions (via Biot-Savart induction) on fluid elements, hence on particles.
It is reasonable to assume that for a given separation $l$ the relative velocity induced by the larger scale  turbulent motions CS2 is weak (compared to CS1), albeit non-negligible, and we say that they are interacting non-locally.  
Such processes defines the neighbourhood of non-locality, corresponding to the wavenumber range, $k_1<k\le k_c(t)$ in Fig. \ref{fig_02}(c).  

$k_c$ and $k_l$ are functions of time as the pair separation increases, but the size of the bandwidth of the local neighbourhood relative to the pair separation defined by the ratio,
\begin{eqnarray}\label{eq_2.1}
    R_c &=& \frac{k_l(t)}{k_c(t)} = \frac{l_c(t)}{l(t)}=constant,
\end{eqnarray}
is assumed to be constant, this is consistent with the concept of self-similarity namely the physical processes are similar at all scales.

Let us re-emphasize the difference between local and non-local scaling laws. The local processes can only scale with the parameters $l$, $\varepsilon$, and $t$. The non-local processes scale with an additional outer length scale $L$ which characterizes the large scale turbulence structures.

\begin{figure}
\begin{center}
\begin{subfigure}{0.45\textwidth}
  \includegraphics[width=3.5cm]{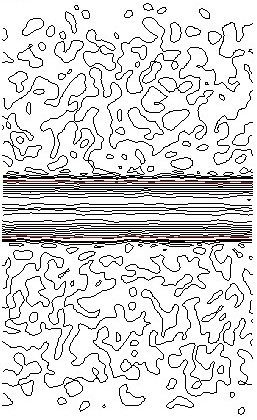}
   \caption{}
\end{subfigure}
\begin{subfigure}{0.45\textwidth}
   \includegraphics[width=3.5cm]{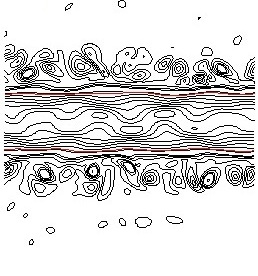}
  \caption{} 
\end{subfigure}\\
\begin{subfigure}{0.45\textwidth}
  \vspace{1cm}
  \includegraphics[width=3.5cm]{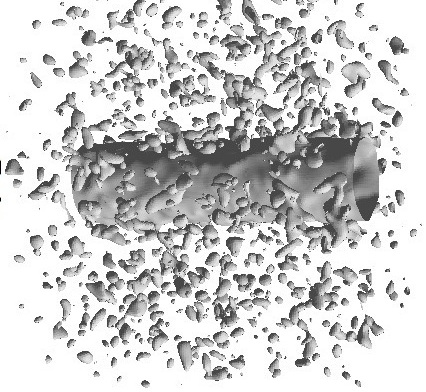}
   \caption{}
\end{subfigure}
\begin{subfigure}{0.45\textwidth}
   \includegraphics[width=4.06cm]{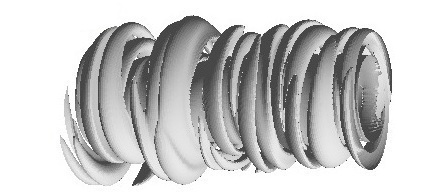}
  \caption{} 
\end{subfigure}\\
   \caption{(Colour online). {DNS results  (Re=5000), 2\% turbulence intensity) from Hussain \& Stout (2013) of dynamics of the growth on a vortex core set in an initial small scale random turbulent field. Top: Vorticity magnitude contours  in the meridional (x-z) plane with different contour levels (a) 0, (b) 40. Bottom: Vorticity magnitude surface at (c) 0, (d) 40. (a) and (c) are at the initial time, and (b) and (d) are at 40 turnover time.}}    \label{fig_03}
\end{center}
\end{figure}

An example of the existence and interaction of  local and non-local scales of motion can be seen in the development of a vortex core in an initially random turbulent field, as seen in Fig. \ref{fig_03} from the the DNS study by \cite{Hussain2013}. Bio-Savart induction from the core organizes the small scale random turbulence outside of the core in to larger organized vortex rings wrapping around the vortex core. These rings evolve as pairs which annihilate each other by cross-diffusion leaving monopole threads which induce organized long wavelength Kelvin waves within the core. This therefore is an excellent example of feedback between the large and small scales of turbulence.
  
Another useful analogy is the turbulent  boundary layer where both short range interactions in the near wall region and long range interactions in the outer turbulent layer exist, illustrated in Fig. \ref{fig_04}.  Long range energetic motions induce flow in the near wall region (Biot-Savart law) leading to the formation and development of hairpin structures. The boundary layer grows outward downstream and interact with the larger outer structures thereby providing  a type of feedback to the upper boundary layer. These highly complex non-linear processes again illustrate the existence of long-range and short-range interactions acting in the same hydrodynamic system which are analogous to the pair diffusion processes.

\begin{figure}
\begin{center}
   \centerline{\includegraphics[width=13cm]{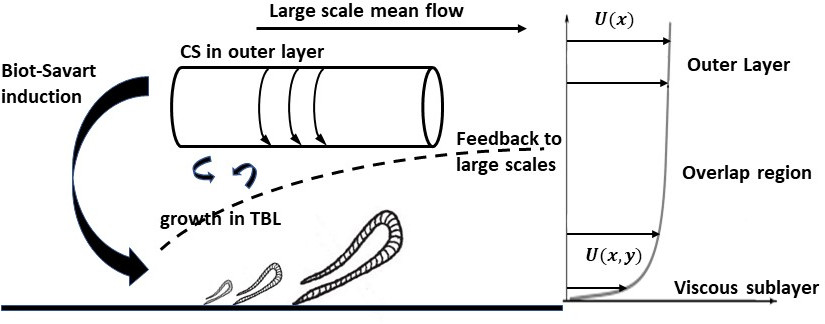}}
   \caption{Sketch of a typical boundary layer illustrating different turbulent motions indicated by the arrows in the near wall and outer regions. Possible interactions between large scale flow structures (CS) and the small scales of turbulent motions in the TBL are indicated. The passage of large scales induces thin shear layers near the wall via Biot-Savart induction and generate near wall structures via transient growth (Schoppa \& Hussain 2002). The turbulence structures in the TBL grow downstream and interact with the larger outer scales thus completing the feedback. The mean streamwise velocity profile is shown to the right.}    \label{fig_04}
\end{center}
\end{figure}

%%%%%%%%%%%%%%%%
\section{Non-local scaling laws for pair diffusivity}\label{SS2.2}

\subsection{The theory}

We stress that the partitioning of the integral for the diffusivity into contiguous neighbourhoods in equation (\ref{eq_15})  is an exact mathematical operation -- it is {\em not} an approximation or an ad hoc assumption. It means that {\em a priori} we cannot dismiss the existence of either the local or the non-local processes. Thus, it is the locality in space hypothesis (which is implicitly assumed in the R-O theory) which turns out to be an ad hoc assumption, not the theory being advanced here.

The integral defining $K^{nl}$ in equation (\ref{eq_15}) is taken over scales of motion much larger than $l$, i.e. in a range of wavenumbers, $k_1\le k\le k_c\  (< k_l)$. Physically, the diffusional process in this neighbourhood is deemed to scale non-locally to $l$, illustrated in Figs. \ref{fig_02}-\ref{fig_04}. Mathematically, within this non-local neighbourhood of  we have $|{\bf k\cdot l}|\ll 1$ and the integrand in (\ref{eq_15}) can be approximated accurately by, $\exp\left({ i{\bf k\cdot l}}\right)-1 \approx i{\bf k\cdot l}$. 

However, the integral defining $K^l$ in equation (\ref{eq_15}) is taken over scales of motion that are close to $l$, i.e. in a local range of wavenumbers,  $k_c\le k\le k_l$.  Physically, the diffusional process here scales locally to $l$, illustrated  in Fig. \ref{fig_02}(a) by the pair A-B which are inside the same typical structure. Mathematically, in the integrand for $K^l$ an expansion of the exponential term, $\exp\left({ i{\bf k\cdot l}}\right)-1 \approx i{\bf k\cdot l}$ to leading order is not completely accurate because $|{\bf k\cdot l}|\approx 1$. Nevertheless, we can still assume this to be approximately true but with a correction factor $F_l<1$ to the overall integral (which accounts for the extra terms in the expansion of the exponential).

Assuming the closure $ \langle ({\bf l\cdot A}) ({\bf k\cdot l})\rangle \sim  l^2 \langle kA\rangle$, and integrating over all directions, we obtain from equation  (\ref{eq_15})
\begin{eqnarray}\label{eq_23}
	K(l) &\sim&  l^2 \int_{nl} \sqrt{kE(k)} dk } +  F_l {\int_{l} l^2 \sqrt{kE(k)} dk.
\end{eqnarray}
In the first term, non-locality means that we take $ l^2$ outside of the integral because the pair separation $ l^2$ is only weakly coupled to the turbulent motions CS2 in the non-local neighbourhood. However, in the local integral on the right the pair relative motion is strongly coupled to the local flow structure CS1 in Fig. \ref{fig_02}(a), and we apply the local scaling $ l^2 \sim k^{-2}$ inside the integral. This leads to the following expressions for the non-local term,  assuming the spectrum in equation (\ref{eq_2.2}),
\begin{eqnarray}\label{eq_34}
	K^{nl} (l) &\sim& \varepsilon^{1/3} L^{-2/3} l^{\gamma^{nl}},    
     \qquad {\rm with} \quad  \gamma^{nl}(p)=2, \quad    1<p\le 3
\end{eqnarray}
$\gamma^{nl}(p)$ is the non-locality scaling, and it is always equal to $2$, independent of $p$. For the local term we obtain,
\begin{eqnarray}\label{eq_30}
	K^l (l)  &\sim& F_l \varepsilon^{1/3} L^{(5/3-p)/2} l^{\gamma^l }, 
     \quad{\rm where}\quad   \gamma^l(p) = (1+p)/2, \quad  1<p\le 3
\end{eqnarray}
For Kolmogorov turbulence, $p=5/3$, this gives, $K^l\sim l^{4/3}$. In the R-O theory, it is assumed that $k^{nl}=0$ which recovers the 4/3-scaling law, $K\approx  K^l\sim l^{4/3}$.

In the new theory, the non-local and local diffusional processes are deemed to be approximately independent, i.e. we ignore the effect of CS2 on CS1, leading to
\begin{eqnarray}\label{eq_2.4}
     K(l) &\sim&  O\left(l^{2}\right) + O\left(l^{(1+p)/2}\right).
\end{eqnarray}

Richardson assumed that there was a single unique power law over the entire inertial subrange. This can be justified on the principle of self-similarity of the diffusional processes at different scales in the system, equation (\ref{eq_2.1}), from which it follows that there can only be a unique power scaling across the entire inertial range if it is of infinite extent, $R_k\to\infty$, or at least in a range of separations that is well inside a finite but large inertial subrange $1\ll l/\eta \ll R_k<\infty$ which ensures that the ultra-violet (UV) and Infra-Red (IR) corrections from the ends of the inertial subrange are negligible. Assuming that the conditions for observing inertial subrange scaling are met, then applying this principle to equation (\ref{eq_2.4}) yields, 
\begin{eqnarray}\label{eq_2.5}
     K(l) \sim l^{\gamma}, \quad {\rm with} \qquad  \gamma^l(p) \le \gamma (p)\le \gamma^{nl}(p). 
\end{eqnarray}
The prediction for the scaling exponent is intermediate between the purely local $\gamma^l$  and purely non-local $\gamma^{nl}$.

However, the scaling law is sensitive to the balance of the non-local and local terms defined as $M_K=K^{nl}/K^l$ which is a function of $p$,  and $R_l$, and $R_c$ which is assumed to be a constant (for a given $p$), equation (\ref{eq_2.1}) and see Fig. \ref{fig_02}(c). $M_K$ is obtained by dividing equation (\ref{eq_34}) by equation (\ref{eq_30}) and after some further algebra,
\begin{eqnarray}\label{eq_3.6b}
   M_K &=& \frac{1-\displaystyle{\left(\frac{R_c}{R_l}\right)^{(3-p)/2}}}{F_l \left({ R_c^{(3-p)/2} -1}\right)},
                   \quad {\rm for}\  R_l\ge R_c. 
\end{eqnarray}

For spectra close to $p=5/3$, and for separations well inside the inertial subrange $1\ll l/\eta \ll R_k$, $M_k$ displays two limiting cases. 

First, as the inertial subrange becomes small as $k_c\to k_1$, so that $R_l\downarrow R_c$  in Fig. \ref{fig_02}(c)),  the non-local neighbourhood shrinks and disappears (because $k_c/k_1=1$)  leaving just the local neighbourhood. In this limit $M_k\to 0$ because the non-local diffusional process has less and less influence, and $K$ asymptotes towards approximate (quasi-) local scaling laws, with $\gamma \approx \gamma^l = (1+p)/2$,
\begin{eqnarray}\label{eq_2.6}
    K &\to& l^{(1+p)/2},\quad {\rm as}\quad R_k\downarrow R_c.
\end{eqnarray}
In this limit the entire inertial subrange, $k_1=k_c\le k\le k_\eta$,  is the local neighbourhood to the pair separation, Fig. \ref{fig_02}(c).
(This result also proves the existence of the local neighbourhood. The existence of the non-local neighbourhood can also be established explicitly, see Appendix B.)
The important distinction between Richardson's locality concept and this new quasi-local limit will be elaborated on later after the numerical results are presented.

\begin{figure}
\begin{center}
\begin{subfigure}{0.48\textwidth}
  \includegraphics[width=6.5cm]{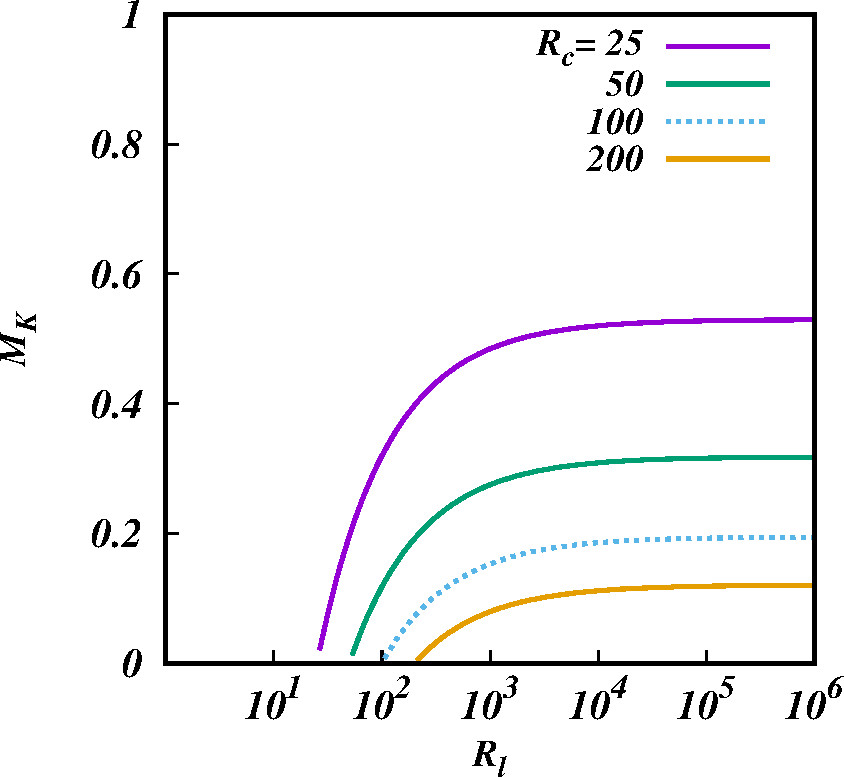}
   \caption{}
\end{subfigure}
\begin{subfigure}{0.48\textwidth}
   \includegraphics[width=6.5cm]{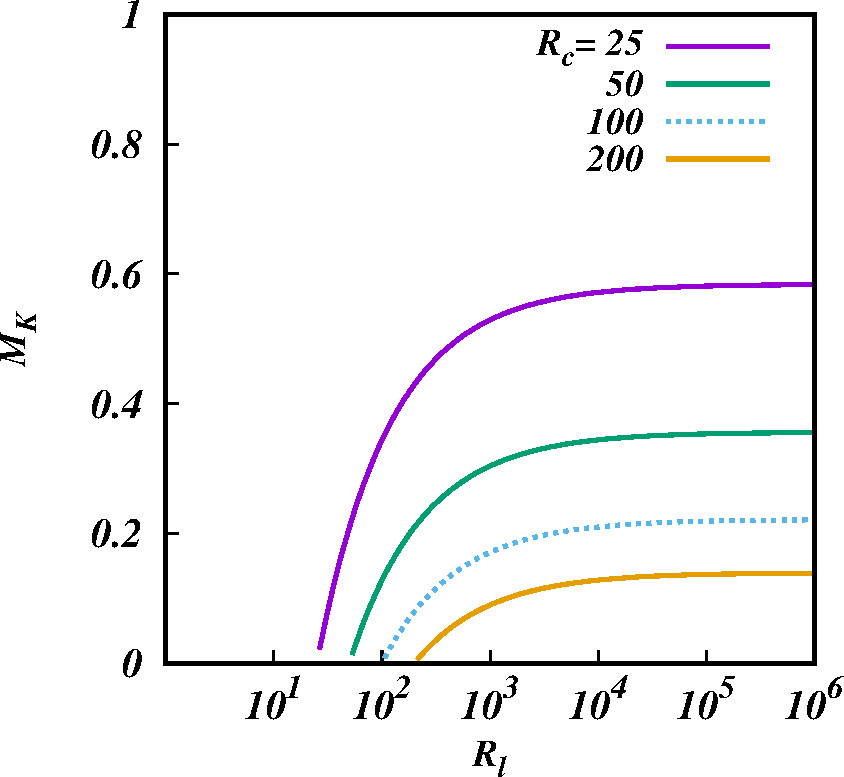}
  \caption{} 
\end{subfigure}\\
   \caption{(Colour online). {$M_k$ (equation \ref{eq_3.6b}) against $R_l$ for different $R_c$ as indicated. For spectra, (a) $E(k)\sim k^{-5/3}$ and (b) $E(k)\sim k^{-1.72}$.}}    \label{fig_05}
\end{center}
\end{figure}

\begin{figure}
\begin{center}
\begin{subfigure}{0.48\textwidth}
  \includegraphics[width=5.5cm]{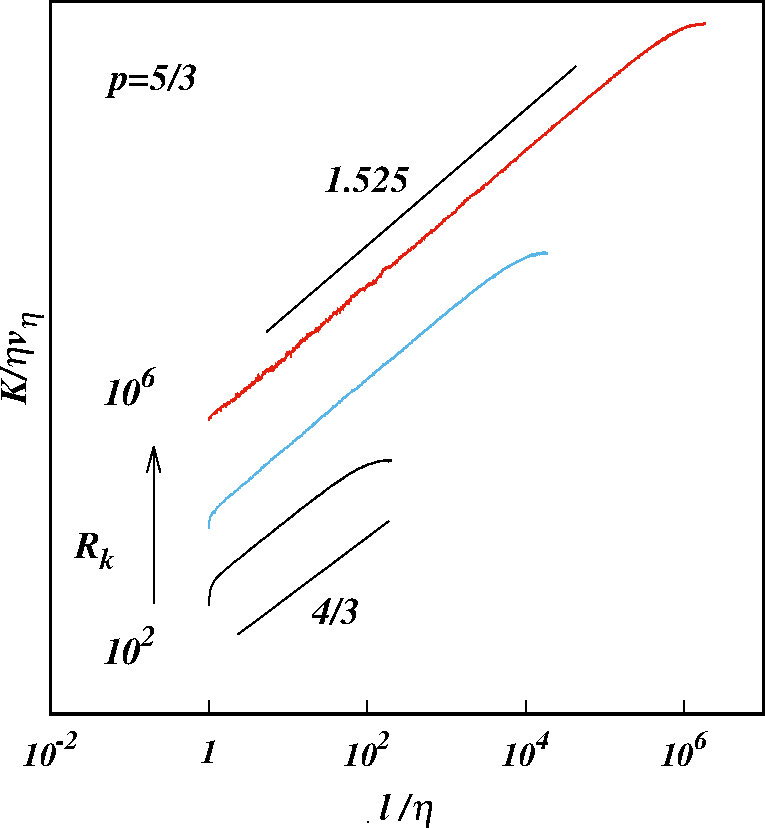}
   \caption{} 
\end{subfigure}
\begin{subfigure}{0.48\textwidth}
   \includegraphics[width=5.5cm]{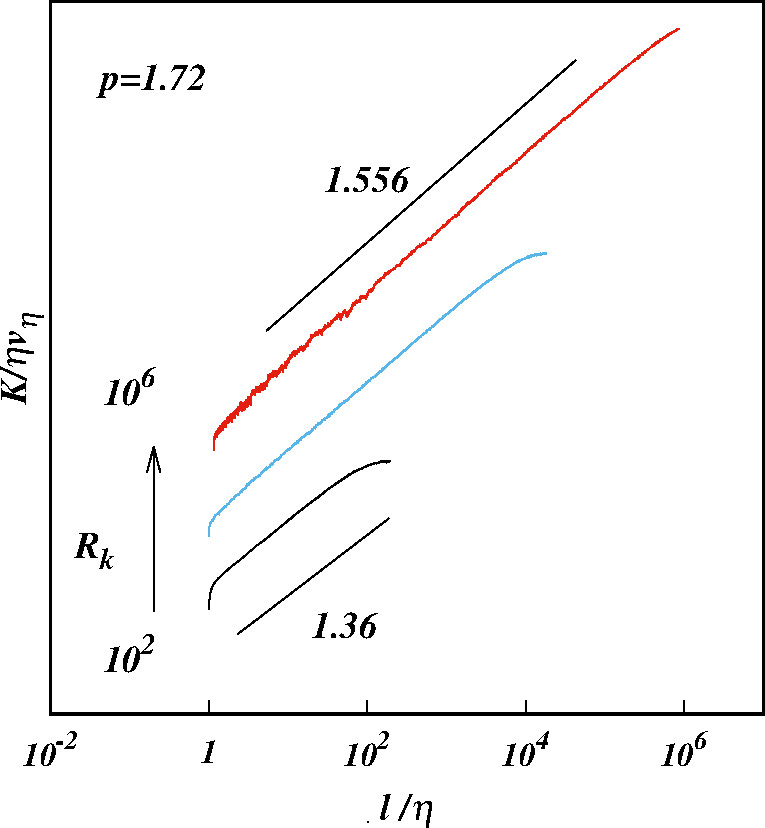}
  \caption{}
\end{subfigure}\\
   \caption{(Colour online). {$K/\eta v_\eta$, against $l/\eta$, for $R_k=10^2$, $10^4$, and $10^6$ as indicated. From simulations with energy spectra, (a) $E(k)\sim k^{-5/3}$ and (b) $E(k)\sim k^{-1.72}$. Lines of various slopes are shown for comparison.}}    \label{fig_06}
\end{center}
\end{figure}

\begin{figure}
\begin{center}
   \includegraphics[width=7cm]{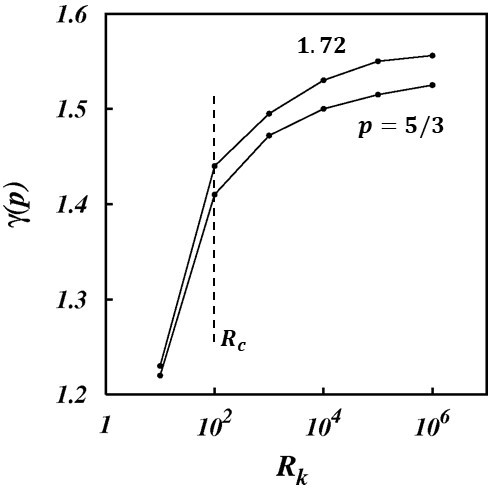}
   \caption{The scaling law $\gamma(p)$ against $R_k$ from the simulations with spectra $E(k)\sim k^{-5/3}$ and $\sim k^{-1.72}$. The size of the locality bandwidth, $R_c=10^2$, is indicated by the dashed line. For $R_k<R_c$ IR corrections dominate and we do not see inertial subrange scaling at all.}     \label{fig_07} 
\end{center}
\end{figure}

Second, in the limit of infinite inertial subrange $R_k > R_l\to\infty$ we obtain $M_K\approx R_C^{-2/3}/F_l \approx O(1)$ for realistic values of $F_l$ and $R_c$ (see below). 

However, $M_K$ is sensitive to the value of  the locality bandwidth $R_c$; in general as $R_c$ increases $\to R_l$ , then $M_K$ becomes smaller because the relative importance of the non-local effects decreases as can be seen from Fig. \ref{fig_02}(c).
We illustrate by considering two particular spectra, (a) $E(k)\sim k^{-5/3}$, and (b) $E(k)\sim k^{-1.72}$, the latter coming from turbulence with intermittency correction (measured experimentally the range of accepted values is $1.692\ <p<1.741$ \cite{Anselment2001, Tsuji2004, Tsuji2009, Meyers2008}). Fig. \ref{fig_05} shows $M_K$ against $R_l$ from equation (\ref{eq_3.6b}) for $R_c=25, 50,100,200$, and $F_l=0.25$. We observe that the local diffusional processes are always important, whereas the non-local diffusional processes only become important as the size of the inertial subrange increases although they are never dominant with an upper bound of about $M_K\lesssim 0.3$.  The steeper intermittent spectrum shows slightly higher $M_K$, Fig. \ref{fig_05}(b), reflecting a small shift towards non-locality.

\subsection{Simulation results for pair diffusion}

Ideally, we would like to investigate the theory through Direct Numerical Simulations DNS. However, DNS cannot generate the very high Reynolds number, or the size of the inertial subrange required for this study, so we have to turn to diffusion models. Most diffusion models are formulated in the Lagrangian frame of reference and therefore are kinematic in nature in the sense that they do not solve the dynamical equations of motions, rather they pose models for the particle transport. Stochastic models for example pose probability distributions for the  incremental particle motions (velocities, accelerations) from timestep to timestep. 
In this study, we will use Kinematic Simulations (KS) because it is the only diffusion model at the current time that can generate the large inertial ranges needed to investigate asymptotically infinite Reynolds number turbulence, and it contains the correct energy spectrum of turbulence, and it is incompressible by construction.  Previous criticism of KS that it does not adequately account for the sweeping of small scales of motion by the larger scales of motion have been disproved, details of which and a summary of the KS method is contained in Appendix A. 

The simulations were run for inertial subrange sizes in the range, ${10}^1\le R_k\le{10}^6$.  The frequencies are defined with $\lambda=0.5$ (Appendix A) which is the standard case in most KS studies. We take a large ensemble of about 30,000 particle pairs, and a very small timestep; the statistical errors in all the results are negligible.

Computed values of $\gamma$ and other quantities for the spectra $E(k)\sim k^{-5/3}$, and $E(k)\sim k^{-1.72}$ are shown in Tables 1 and 2 respectively.  Fig. \ref{fig_06}(a) shows log-log plots of the pair diffusion coefficients, $K/\eta v_\eta$, against  $l/\eta$ for different sizes of the inertial subrange, $R_k$, from simulations with Kolmogorov spectrum $E(k)\sim k^{-5/3}$ -- only the cases with $R_k=10^2, 10^4, 10^6$ are shown for clarity. We observe clear power-laws as predicted in equation (3.5) across most of the inertial subrange. Fig. \ref{fig_06}(b) shows the results from turbulence with intermittency correction, and we see similar trends as in Fig. \ref{fig_06}(a).

Fig. \ref{fig_07} shows the scaling laws $\gamma$ against the size of the inertial subrange $R_k$ obtained from the simulations for the same two spectra, $p=5/3$ and $1.72$. There is a smooth transition in $\gamma$ from the quasi-local regimes at $R_k=10^2$ to the asymptotic near infinite regimes at $R_k=10^6$. For $R_k<10^2$ we loose inertial subrange scaling altogther presumably due to IR corrections from the upper end of the inertial subrange.

For the spectrum $E\sim k^{-5/3}$, we obtain $\gamma\approx 1.41$ at $R_k=10^2$ (which  is about $6\%$ greater than the pure local scaling of $4/3$), and $\gamma=1.525$ at $R_k=10^6$,
\begin{eqnarray}
   K\sim l^{1.41},\quad   for\quad R_k\approx{10}^2    \label{eq_2.7}\\
   K\sim l^{1.525},\quad for\quad R_k\approx{10}^6.   \label{eq_2.8}
\end{eqnarray}
For the spectrum $E\sim k^{-5/3}$, we obtain $\gamma\approx 1.44$ at $R_k=10^2$ (which  is about $8\%$ greater than the pure local scaling), and $\gamma=1.556$ at $R_k=10^6$,
\begin{eqnarray}
   K &\sim& l^{1.44},   \quad for\quad R_k\approx{10}^2      \label{eq_2.9}\\
   K &\sim& l^{1.556}, \quad for\quad R_k\approx{10}^6      \label{eq_2.10} 
\end{eqnarray}
}

\begin{table*}
\begin{center}
\def~{\hphantom{0}}
{
}
\begin{tabular}{lll ll}
\hline \\
$R_k$	   &$\gamma$    &$\chi$   &$G_K$	   &$G_l$ \\		
\hline \\
$10^1$	&1.220  &1.89   &1.20  &1.25         \\[-2pt]
$10^2$	&1.410  &2.94   &0.82  &1.36E-1   \\[-2pt]
$10^3$	&1.472  &3.58   &0.74  &3.93E-2   \\[-2pt]
$10^4$	&1.500  &3.90   &0.74  &1.99E-2   \\[-2pt]
$10^5$	&1.515  &4.11   &0.74  &1.38E-2   \\[-2pt]
$10^6$	&1.525  &4.20   &0.74  &1.29E-2   \\[-2pt] \\
\hline
\end{tabular}
 \caption{{ From the simulations with $E(k)\sim k^{-5/3}$; $L=1$. The calculated quantities are defined in the text.}\protect}\label{Table2}
\end{center}
\end{table*}

\begin{table*}
\begin{center}
\def~{\hphantom{0}}
{
}
\begin{tabular}{lll ll}
\hline \\
$R_k$	  &$\gamma$   &$\chi$    &$G_K$  &$G_l$ 	\\
\hline \\
$10^1$	 &1.23    &1.88   	&1.17    &1.22       \\[-2pt]
$10^2$	 &1.44    &3.00   	&0.79    &1.16E-1  \\[-2pt]
$10^3$	 &1.495  &3.70  	&0.73    &2.54E-2  \\[-2pt]
$10^4$	 &1.530  &4.11  	&0.73    &1.07E-2  \\[-2pt]
$10^5$	 &1.550  &4.38  	&0.73    &6.70E-3  \\[-2pt]
$10^6$	 &1.556  &4.50       &0.73     &6.10E-3  \\[-2pt] \\
\hline
\end{tabular}
 \caption{{ From the simulations with $E(k)\sim k^{-1.72}$; $L=1$. The calculated quantities are defined in the text.}\protect}\label{Table3}
 \end{center}
 \end{table*}

Equation (\ref{eq_2.10}) is especially important because this scaling law is within 1\% of the re-appraised Richardson's 1926 dataset from geophysical turbulence measurements which displays, $K\sim l^{1.564}$. As the 1926 dataset remains the only fully 3D dataset of pair diffusivity on a geophysical scale, i.e. in large scale 3D turbulence containing effectively infinite inertial subrange, the simulation result (\ref{eq_2.10}) provides some degree of verification for the non-local theory.

To understand Infra-Red (IR) corrections in the limit $R_k\downarrow1$, consider what happens as the pair separation increases in the limit $l\uparrow L_1$. Then, the influence from the very large scales larger than $L_1$ must be considered because they induce relative motions at higher separations of sufficient magnitude that may notably alter the pair diffusion as  $l\uparrow L_1$, especially in short inertial subranges. Off course, for separations $l\ll L_1$ in large inertial subranges (in the limit of $R_k\to\infty$) this effect is negligible because such large scales ‘sweep’ the pair along, as noted earlier. 

Equations (\ref{eq_2.7}) and (\ref{eq_2.9}) are important because $R_k\approx 10^2$ is the minimum inertial subrange which displays clear quasi-local scaling laws in $K$. For smaller $R_k<10^2$, Infra-Red (IR) corrections in particular from the very large scales dominate and inertial subrange scaling disappears all together. Importantly, we can estimate a value for $R_c$ by equating it  to this minimum of $R_k$,  yielding $R_c\approx 10^2$. From Fig. \ref{fig_05}, the cases for $R_c=50$  and $100$ are probably the  most realistic,  so $R_c\approx 10^2$ as an order of magnitude is a reasonable estimate.

%%%%%%%%%%%%%%%%
\section{\textbf Estimates for $g_l$, and new constants $G_K$ and $G_l$}\label{Sec3}

\subsection{\textbf Why is $g_l$ not a constant?}\label{Sec3.1}

\begin{table*}
\begin{center}
\def~{\hphantom{0}}
{\linespread{1.2}
}
{\begin{tabular}{@{}lll l} \hline \\
& \# & \multicolumn{1}{c}{$g_l$} & References  \\ \hline \\ [0.5ex]
&1    & 0.06   & \cite{Tatarski1960} \\ % ,Sawford2001} \\ 
&2    & 2.42   & \cite{Kraichnan1966} \\ % ,Sawford2001} \\ 
&3    & 0.5     & \cite{Julian1977} \\ % ,Salazar2009}\\ 
&4    & 2.34   & \cite{Lundgren1981} \\ % ,Sawford2001} \\
&5    & 3.52   & \cite{Lesieur1981,Larcheveque1981} \\ % ,Salazar2009} \\ 
&6   & 0.9     & \cite{Borgas1994} \\ % ,Sawford2001}\\ 
&7   & 0.5     & \cite{Borgas1994} \\ % ,Sawford2001}\\ 
&8   & 0.8     & \cite{Borgas1994} \\ % ,Sawford2001}\\ 
&9    & 0.35   & \cite{Pedrizzetti1994,Heppe1998} \\ % ,Sawford2001} \\ 
&10  & 2.0     & \cite{Kurbanmuradov1995} \\ % ,Sawford2001} \\ 
&11  & 1.4     & \cite{Thomson1996} \\ % ,Sawford2001} \\ 
&12  & 0.06   & \cite{Elliott1996} \\ % , Sawford2001}  \\ 
&13  & 1.0     & \cite{Borgas1998} \\ % ,Sawford2001}\\ 
&14  & 2.0     & \cite{Borgas1998} \\ % ,Sawford2001}\\ 
&15  & 0.5     & \cite{Mann1999} \\ % ,Sawford2001} \\ 
&16  & 0.55   & \cite{Boffetta2002} \\ % ,Salazar2009}\\ 
&17  & 0.7     & \cite{Ishihara2002} \\ % ,Salazar2009}\\ 
&18  & 0.83   & \cite{Yeung2004} \\ % ,Salazar2009}\\
&19  & 0.5     & \cite{Biferale2005} \\ % ,Salazar2009}\\
&20  & 0.55   & \cite{Berg2006} \\ % ,Salazar2009}\\ 
&21  & 1.15   & \cite{Berg2006} \\ % ,Salazar2009}\\ 
&22  & 0.5     & \cite{Franzese2007} \\ % ,Salazar2009} \\ 
&23  & 0.57   & \cite{Sawford2008} \\ % ,Salazar2009}\\ 
&24  &0.1      & \cite{Ni2013}\\
&25  &0.55    & \cite{Buaria2015}\\
&26  &0.6      & \cite{Darragh2020}\\
\hline 
\end{tabular}} 
 \caption{The estimates for the Richardson-Obukhov constant $g_l$ from experiments (numbered 1, 3, 15, 20, 21, and 24) and simulations (excluding KS) --   details of methods are in the references cited. The estimates are listed in order of year of estimate. \protect} \label{Table1}
\end{center}
\end{table*}

\begin{figure}
\begin{center}
   \centerline{\includegraphics[width=8cm]{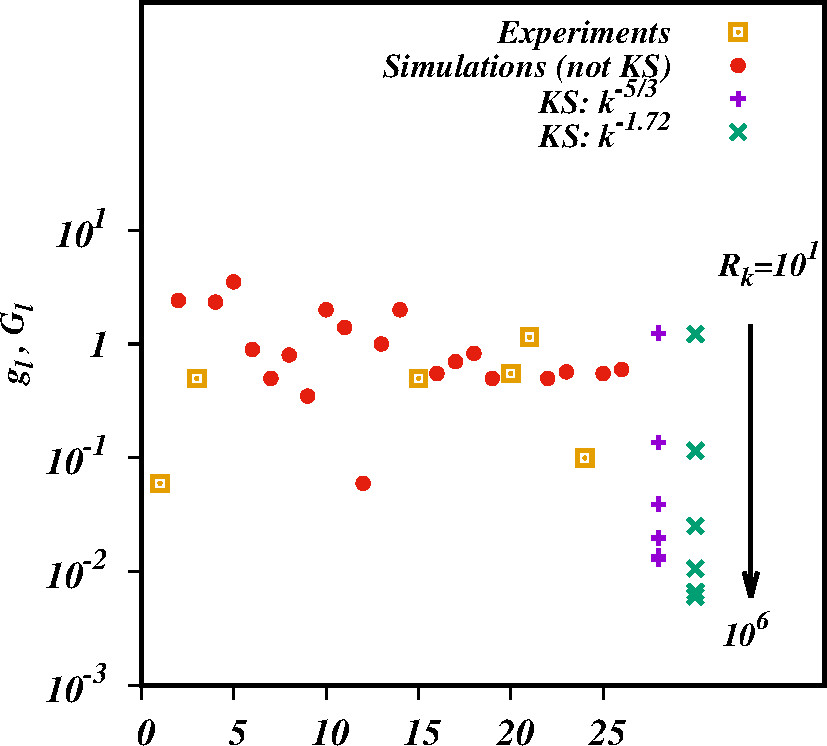}}
   \caption{(Colour online). {$g_l$ from experiments ($\square$ squares) and simulations (not KS, $\circ$ circles), both listed in Table \ref{Table1} in order of year of estimate  (absicca); and $G_l$ from the simulations for different size of the inertial subrange from $R_k=10^1$ to $10^6$ as indicated by the arrow, with $E(k)\sim k^{-5/3}$ (+ symbols, Table 2), and $E(k)\sim k^{-1.72}$ ($\times$ symbols, Table 3).}}    \label{fig_08}
\end{center}
\end{figure}

Richardson's locality in space hypothesis leads to the well-known 4/3-scaling for the turbulent pair diffusivity inside an infinite inertial subrange,
\begin{eqnarray}\label{eq_3.1}
     K(l) &=& g_K\varepsilon^{1/3}l^{4/3},\quad R_k\to \infty  
\end{eqnarray}
This is equivalent to (\cite{Obukhov1941}),
\begin{eqnarray}\label{eq_3.2}
     l^2  &=& l_0^2 + g_l \varepsilon t^3, \quad R_k\to \infty
\end{eqnarray}
$g_l$ is assumed to be a universal constant. $g_K$ is related to $g_l$ and is also a universal constant, but historically all the attention has been focused on $g_l$.

However, in the non-local theory in which both local and non-local processes govern turbulent pair diffusion (Section 3) new non-local power law regimes emerge, 
\begin{eqnarray}\label{eq_3.3}
    K(l) &=& G_KL^{4/3\ -\gamma}\varepsilon^{1/3}l^\gamma, \quad   R_k\to \infty
\end{eqnarray}
or, non-dimensionalizing with $\eta$, $t_\eta$ and $v_\eta\sim \eta/t_\eta$, we obtain
\begin{eqnarray}\label{eq_4.2}
    \frac{K}{\eta v_\eta} &=& G_K \left({\frac{L}{\eta}}\right)^{4/3\ -\gamma} 
                                                       \left({\frac{\eta\varepsilon}{v_\eta^3}}\right)^{1/3} 
                                                       \left({\frac{l}{\eta}}\right)^\gamma,   \quad      0\ll l/\eta\ll R_k\to \infty.
\end{eqnarray}
where $(1+p)/2<\gamma(p)<2$.  For the pair separation, we have
\begin{eqnarray}\label{eq_3.5}
    l^2  &=& l_0^2 + G_l L^{2-2\chi/3} \varepsilon^{\chi/3} t^\chi, \quad       R_k\to \infty
\end{eqnarray}
or in non-dimensional form,
\begin{eqnarray}\label{eq_4.3}
   \frac{ l^2}{\eta^2}  &=&  \frac{l_0^2}{\eta^2}   + 
                                        G_l  \left({\frac{L}{\eta}}\right)^{{2-2\chi/3}}   
                                        \left({\frac{\eta\varepsilon}{v_\eta^3}}\right)^{\chi/3} 
                                        \left({\frac{t}{t_\eta}}\right)^{\chi},  \quad  0\ll l/\eta\ll R_k\to \infty.
\end{eqnarray}
(R-O locality scaling laws, equations (\ref{eq_3.1}) and (\ref{eq_3.2}), corresponds to $\gamma=4/3$.) We also have,
\begin{eqnarray}
   G_l   &=&   \left(\frac{{2G}_K}{\chi}\right)^\chi      \label{eq_3.6}\\
   G_K  &=&  \frac{\chi G_l^{1/\chi}}{2}\ \ \                \label{eq_3.7}\\
   \chi   &=&  \frac{2}{2-\gamma}\ \ \                           \label{eq_3.8}
\end{eqnarray}

A new external length scale $L$ is essential for dimensional consistency, as mentioned earlier. $L$ is some large scale characterizing the turbulence, and it is convenient here to assume that it scales with the largest length-scale in the inertial subrange, $L\sim L_1= 1/k_1$ (although other choices are possible). If we scale all lengths by $L_1$, then we can proceed by taking $L_1=1$.
$G_K$ and $G_l$ are related constants through the non-linear relationship in equation (\ref{eq_3.6}). In the locality limit $\chi\to  3$ , we obtain, $G_K\to  g_K$, and  $G_l\to  g_l =\left({\frac{2g_K}{3}}\right)^3$.

Note that $g_K$ and  $G_K$ are not direct measures of the pair diffusivity, and $g_l$ and $G_l$ are not direct measures of the spread of particles -- only $K(l)$ and $l^2$ themselves tells us these things. 

Historically, researchers have focused attention exclusively on $g_l$, and estimates for $g_K$ are therefore not available. A summary of estimates for $g_l$  is listed in Table \ref{Table1} in order of the year of estimate, and plotted in Fig. \ref{fig_08} in the same order (absicca) and also distinguished either as experiments (squares) or as  simulations (not KS;  circles). There is considerable scatter in the estimated value of $g_l$, differing by nearly two orders of magnitude, the smallest estimate is $0.06$ and the biggest is $3.25$. The average from all 25 measurements listed in Table \ref{Table1} is $g_l\approx1$. 
It is important to note that all of these estimates have been made {\em assuming} a fit to the R-O locality scaling.  Fig. \ref{fig_08} also shows the estimates from KS simulations for different $R_k$ as indicated by the vertical arrow,  also listed in Tables \ref{Table2} and \ref{Table3}.
(Note that an estimate for $g_K$ in the locality limit from equation (\ref{eq_3.7}) with $\chi=3$ is, $g_K=1.5 g^{1/3}_l$, and if we take $g_l\approx 1$, then $g_k\approx 1.5$.)

It was previously assumed that the disparity in the estimates of $g_l$ was probably due to high uncertainties in most experiments (\cite{Sawford2001, Salazar2009}), and because of the short inertial subrange in current DNS $R_k<{10}^2$, which introduces ultra-violet (UV) and infra-red (IR) corrections to the pair diffusivity (discussed below). 

In our non-local theory, $g_K$ and $g_l$ are scale dependent, and an attempt to estimate them by fitting a theoretical locality scaling to the data may indeed lead to a wide range of estimates seen in Fig. \ref{fig_08}. To illustrate this point further, if we try to force equation (\ref{eq_3.5}) as a fit to  equation (\ref{eq_3.2}), then 
\begin{eqnarray}\label{eq_3.10}
      g_l  &=&      G_l L^{2-2\chi/3} \varepsilon^{\chi/3-1} t^{\chi-3}, \quad       R_k\to \infty
\end{eqnarray}
With $\chi\approx 4.5$, we obtain,
\begin{eqnarray}\label{eq_3.11}
      g_l  &\approx & G_l ^{2/3} \left(\frac{l}{L}\right)^{2/3}, \quad       R_k\to \infty
\end{eqnarray}
Thus, $g_l$ depends on both $L$ and the separation $l$.

%%%%%%%%%%%%%%%%

\subsection{\textbf Results for $G_K$ and $G_l$}\label{Sec3.1}

The numerical results are summarized in Tables \ref{Table2} and \ref{Table3} for, respectively, $E(k)\sim k^{-5/3}$ and $E(k)\sim k^{-1.72}$. 

The value of $\chi$ obtained from the simulations converge to the correct theoretical values obtained from equation (\ref{eq_3.8}) as $R_k\to\infty$; but they differ a little for small $R_k$ due to IR and UV corrections, as noted earlier. 
Estimates for $G_K$ and $G_l$ from the simulations are also shown in Tables \ref{Table2} and \ref{Table3}.

\begin{figure}  
\begin{center}
\begin{subfigure}{0.48\textwidth}
   \includegraphics[height=5.5cm]{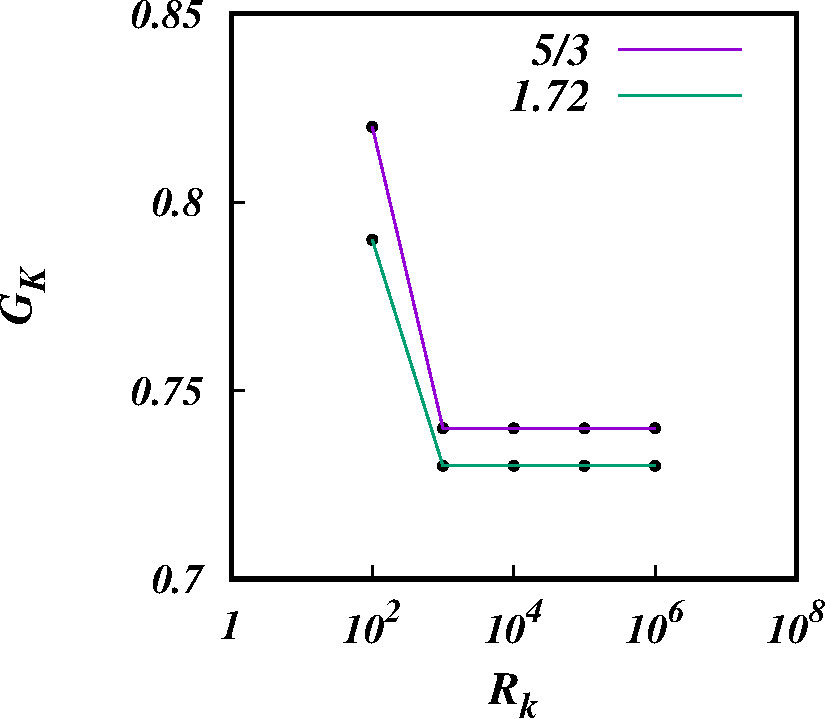}
   \caption{}
\end{subfigure}
\begin{subfigure}{0.48\textwidth}
   \includegraphics[height=5.5cm]{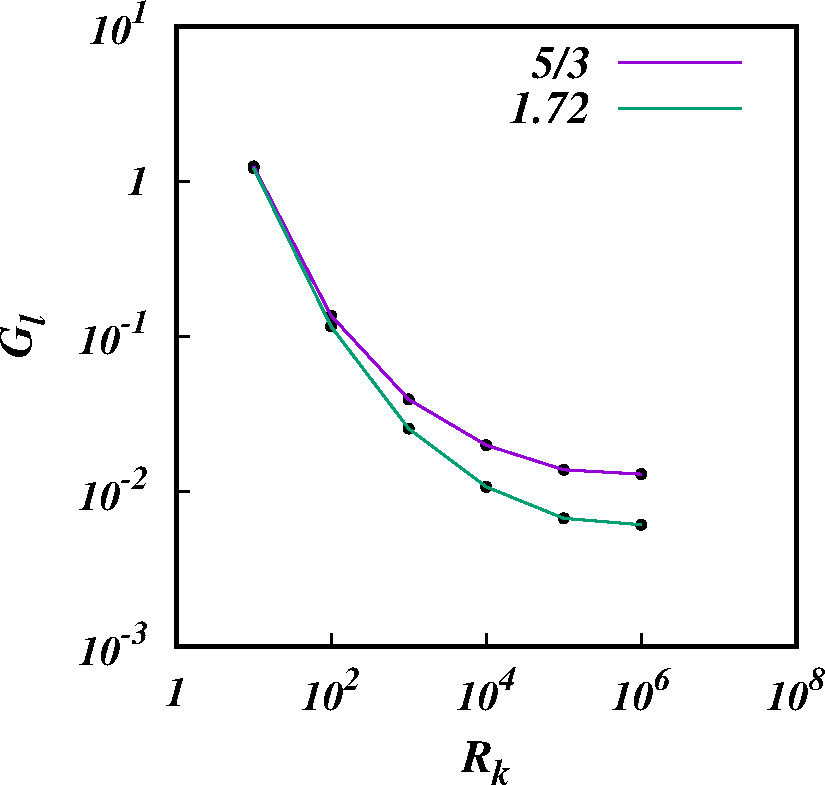}
  \caption{}
\end{subfigure}
  \caption{(Colour online.)  (a) $G_K$ against $R_k$; (b) $G_l$ against $R_k$. From simulations with spectra, $E(k)\sim k^{-5/3}$ and $\sim k^{-1.72}$.} \label{fig_09}
\end{center}
\end{figure}

Fig. \ref{fig_09}(a) shows the plots of $G_K$ against $R_k$ from the spectra $E(k)\sim k^{-5/3}$ and $E\left(k\right)\sim k^{-1.72}$. $G_K$ converges to $G_K\to 0.74$  and $0.73$ respectively for $R_k\geq{10}^3$. As the energy spectrum steepens, there is less energy in the smaller scales, and the decrease in $G_K$ reflects this small shift towards purely non-local diffusion.

Fig. \ref{fig_09}(b) shows the plots of $G_l$ against $R_k$ for the same spectra. As $R_k\to \infty$, the $G_l$ asymptotes to $0.013$ and  $0.006$ respectively. At small $R_k\downarrow10$, the $G_l$'s increase sharply to $O(1)$, which is consistent with most of the numerical and experimental estimates in Table \ref{Table1}. 

It is remarkable how quickly $G_K$ converges from $R_k\ge 10^3$; whereas $G_l$ converges much more slowly as $R_k\to \infty$.
However, these trends are consistent with equation (\ref{eq_3.6}), and the asymptotic limits are given exactly by equation (\ref{eq_3.6}) confirming its validity.

The asymptotic value of $G_l\approx 0.01$ is unexpectedly small. However, it is consistent with some of the experimental results; the two exceptional cases in Table \ref{Table1} with the smallest estimates, $g_l\approx0.06$, come from turbulence with very large inertial subranges, $R_k\to \infty$. \cite{Tatarski1960} obtained data from the trail of comets in the upper atmosphere, while \cite{Elliott1996} simulated a large inertial subrange over 12-15 decades using a Monte Carlo algorithm. In this limit our results  converge to $G_l \approx 0.006$ -- the trend towards a very small value is clear. 

$G_K$ converges from $R_k\ge10^3$ making is an easier quantity to measure than $G_l$. For these reason, it is probably a more reliable quantity to work with than $G_l$ in future studies.

%%%%%%%%%%%%%%%%%
\subsection{Some applications}

A question of great practical interest is whether the non-local picture has a significant effect on predictions of statistical quantities? Certainly, the new diffusion scaling laws can be expected to have an impact on particle spread. Furthermore, a small value of $G_l$ is an indication that the effectiveness of turbulence in spreading particles in the small scales of turbulence is less than expected from R-O locality in which $G_l=g_l\approx O(1)$ at all scales. We therefore attempt to quantify the impact of the non-local theory on particle spread, to an order of magnitude, and briefly discuss some potential areas of application.

Spray technology is important in many industries, particularly for the dispersal of pesticides over crops. During pesticide dispersal using sprayers, only a proportion (typically less than $50\%$ ) of the emitted spray reaches the intended target and the rest is lost through drift and evaporation; furthermore, the drift can be hazardous to neighbouring field crops and for the environment. Knowing the characteristics of droplet dispersal within the turbulent flow around the crop canopy could improve the application efficiency.

\begin{figure}
\begin{center}
   \includegraphics[width=12cm]{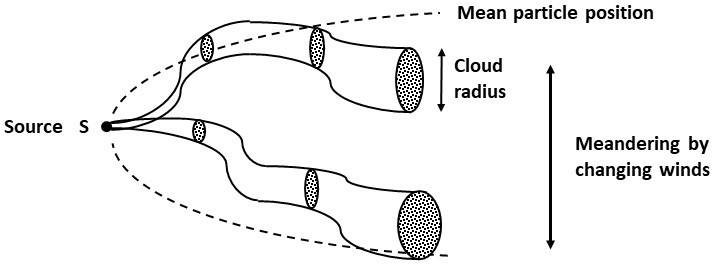}
   \caption{Sketch of the diffusion of a group of particles released from a point source at S (e.g. spray nozzle, chimney) in to the atmosphere. The dashed line is the average position of all particles from all realizations of the atmospheric wind pattern. In an individual release, the instantaneous centroid of the cloud of particles will meander from the overall mean profile, and the cloud radius will grow (like pair diffusion) about the cloud centroid. The centroid pathline will differ in different winds.}
\label{fig_10}
\end{center}
\end{figure}

It is well known that the spread of such contaminant in to the environment has three components, illustrated in Fig. \ref{fig_10}: (i) the average particle position $X_c(t)$ over all realizations; (ii) winds will cause the cloud centroid to move along different meandering paths;
and (iii) the diameter $r_c(t)$ of the cloud of particles, which is related to the particle separation $l(t)$. This system has a been investigated extensively because of its obvious importance to particle spread and pollution modelling, (\cite{Hunt1985,Tampieri2017,Ma2020}). 

The large scale drift is clearly important for the large scale dispersal of the droplets, but this is outside of our control. The small scale turbulent diffusion also plays an important role, because how fast the cloud of droplets spread determines the amount of pesticide that actually reaches the intended target,  and hence the quantity of pesticide to use. Because the canopy height is relatively constant (1-2m in many situations) let us assume that the prevailing weather conditions are constant and then we can ignore variations in the temperature and bouyancy, and we will also ignore evaporation. 

The question is, what is the difference in the spread (radius) of the coud droplets predicted by the current theory as compared to the R-O theory? We have already noted that $M_K\lesssim 0.3$, that is the non-local diffusivity is up to 30\% of the local diffusivity. This might at first glance suggest that the pair diffusivity should increase by about 30\% at any given separation. On the other hand, we have also noted that $G_l\approx 10^{-2}$.   
A comparison of the cloud radius in the non-local and local theories, at the same phase (time after release) is estimated by dividing equations (\ref{eq_3.5}) by (\ref{eq_3.2}) to give,
\begin{eqnarray}\label{eq_4.10}
  M_l &=&   \sqrt{\frac{ l^2} {l^2_{RO}} } 
          \approx \frac{G_l^{1/2}} {g_l^{3/4}}\left({\frac{\eta}{L}}\right)^{1/2} \left({\frac{t}{t_\eta}}\right)^{3/4} 
          \approx 0.2 \left({\frac{\eta}{L}}\right)^{0.5} \left({\frac{t}{t_\eta}}\right)^{0.75},  \eta \ll l \ll L_1.
\end{eqnarray}
if we take $G_l\approx 10^{-2}$, and $g_l\approx 0.5$. The ratio $M_l$ increases with the time after release. For mildly turbulent conditions, i.e. in short inertial subrange, there is little difference. But for stongly turbulent conditions, the factor $\eta/L\ll 1$ ensures that the cloud ratio is smaller than in the R-O theory. To illustrate, in turbulence with $\eta/L=10^4$, the large time scale is $T/t_\eta=\left({L/\eta}\right)^{2/3} =464$, and we have,
\begin{eqnarray}\label{eq_4.11}
      M_l &\approx&  0.06\ \hbox{at}\ t=100 t_\eta \nonumber\\
      M_l &\approx&  0.12\ \hbox{at}\ t=T/2\ \hbox{(deep inside the inertial subrange)} \nonumber \\
      M_l &\approx&  0.20\ \hbox{at}\ t=T\ \hbox{(large time scale)} \nonumber \\
      M_l &\approx&  0.36\ \hbox{at}\ t=1000 t_\eta \approx 2T.
\end{eqnarray}
The cloud radius is reduced by an order of magnitude deep inside a large inertial subrange. 

This will also impact on the large scale drift of the cloud centroid due to the prevailing winds; if the cloud of droplets remains more concentrated for longer periods, then it will drift further in higher concentrations in to neighbouring fields. For example, although the average spread over the crop canopy over all wind conditions ($X_c$) will be unaffected, in individual winds small concentrations of pesticide droplets could drift much further according to the non-local theory.

Another application is in CFD modelling in the small scales through an eddy-viscosity concept; the pair diffusivity itself is a type of eddy-viscosity -- it has the same dimensions. CFD models based on a scale dependent turbulent eddy viscosity such that $\nu_t\sim K\sim l^\gamma$  is a useful avenue to explore in the future.

Recently, \cite{Darragh2020} have explored the variation of pair diffusion statistics in the internal structure of turbulent flames, and they find that the effect of temperature and compressibility is surprising modest and to leading approximation non-reacting turbulent pair diffusion concepts  can be applied directly to combustion processes. Even the R-O `constant' (they consider relatively low turbulence intensity in the quasi-local limit) does not change much with temperature $g_l\approx 0.6$, listed in Table \ref{Table1}.

Although (4.13) are order of magnitude estimates, they are significant because they indicate the possible impact that the non-local theory could have in real world problems.

%%%%%%%%%%%%%%%%%%%%%%%%%%%%%%%%
\section{The impact of changing flow structure}\label{Sec5}

KS is essentially a Fourier series with a finite number of Fourier modes with phase, $\sim e^{-i({\bf k}_n\cdot{\bf x}\ + \omega_n t)}$ (Appendix A). The frequency $\omega_n$ is defined by the physical principle that a trace particle circulates a local vortex like structure in a time inversely proportional to the frequency. That is, the $n'th$ mode frequency is,
\begin{eqnarray}\label{eq_5.1}
  \omega_n &=& \lambda \sqrt{k^3E(k)},
\end{eqnarray}
where $\lambda$ is an arbitrary factor. For Kolmogorov turbulence, $\omega_n=\lambda\varepsilon^{2/3}k_n^{2/3}$.  
The Fourier mode oscillates in wavenumber space manifesting is real space as a periodic velocity field -- $\lambda$ defines the amplitude of the frequency of such an oscillation (this should not be confused with an externally forced oscillation).

Spatio-temporal correlations (manifested as flow structures) are fundamental hallmarks of turbulence. The frequency defines the rate at which CS's change in time, and a natural question is how sensitive  is turbulent diffusion in the inertial subrange to changes in $\lambda$?

Physically, we expect that realistic turbulence is close to $\lambda \approx 1$, and we have already seen that $\lambda = 0.5$ captures the essential scaling laws of turbulent particle diffusion. However, mathematically there is nothing special about this choice, so in the ensuing we will consider a wide range of $\lambda$.

We have run KS  with $E(k)\sim k^{-5/3}$ for $\lambda$ in the range $0\le\lambda\le5$, and for $R_k={10}^2,{10}^4,{10}^6$. The cases for $E(k)\sim k^{-1.72}$ are similar, so we will show results from this spectrum for just one case. 

\begin{figure}
\begin{center}
\begin{subfigure}{0.45\textwidth}
  \includegraphics[width=6cm]{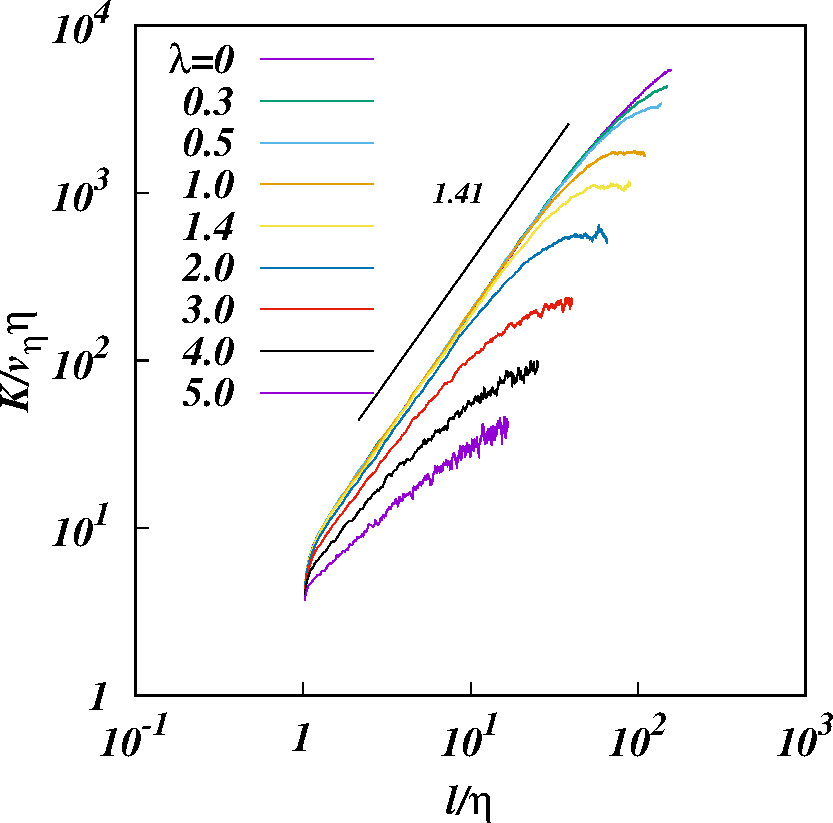}
   \caption{}  
\end{subfigure}
\begin{subfigure}{0.45\textwidth}
   \includegraphics[width=6cm]{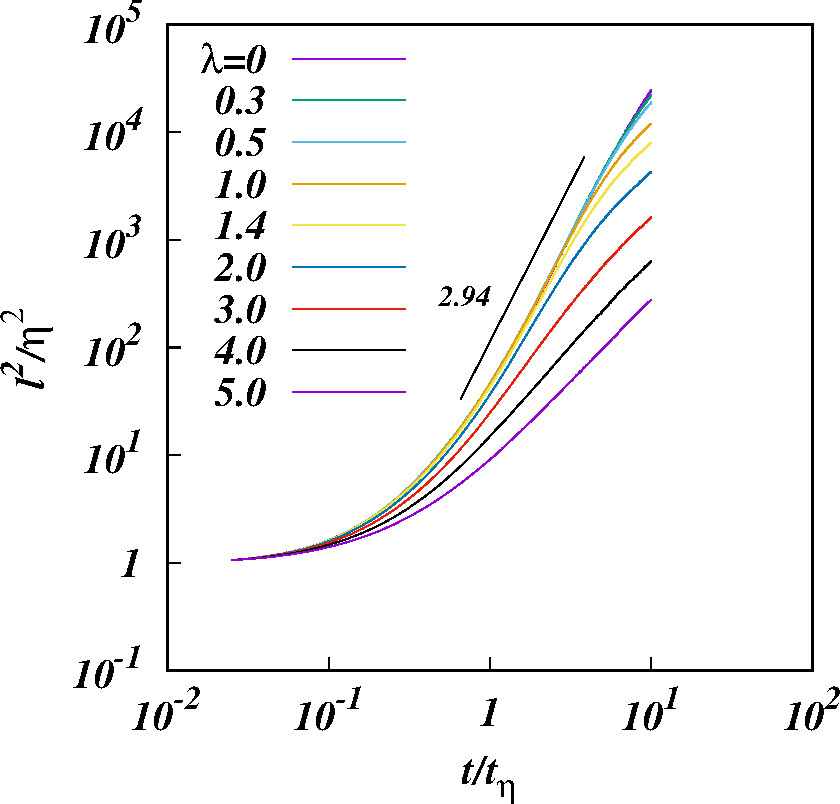}
  \caption{}
\end{subfigure}
\caption{(Colour online.)  From the simulations with spectra $E(k)\sim k^{-5/3}$ and $R_k={10}^2$: (a) $K$ against $l$; (b) $ l^2 $ against $t$. Lines of indicated slope are shown for comparison, see Table 1. Different levels of $\lambda$ are indicated. $\eta$ is the Kolmogorov length scale, $v_\eta$ is the velocity micro-scale, $t_\eta$ is the time micro-scale.} \label{fig_11}
\end{center}
\end{figure}

\begin{figure}
\begin{center}
\begin{subfigure}{0.45\textwidth}
  \includegraphics[width=6cm]{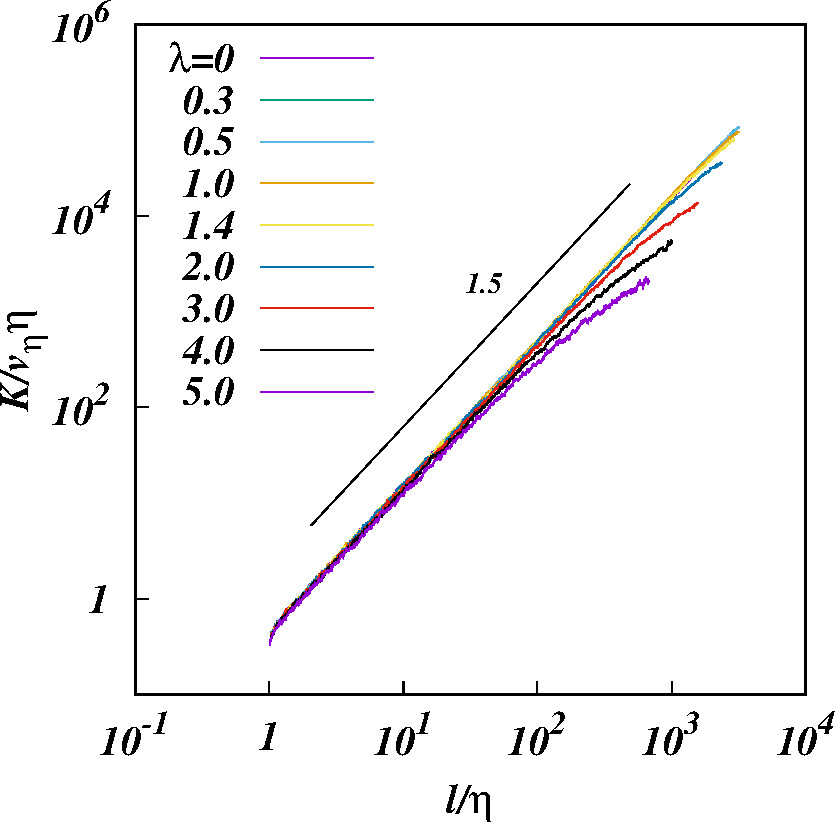}
   \caption{} 
\end{subfigure}
\begin{subfigure}{0.45\textwidth}
   \includegraphics[width=6cm]{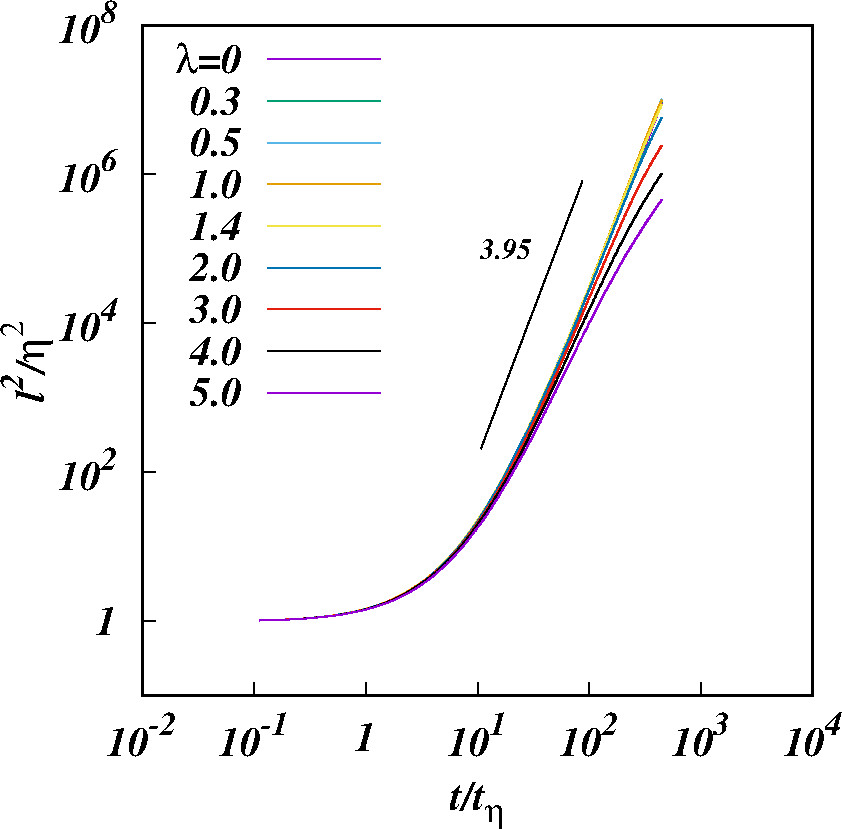}
  \caption{} 
\end{subfigure}
\caption{(Colour online.)  Similar to Fig.  \ref{fig_11}, except for and $R_k={10}^4$. } \label{fig_12}
\end{center}
\end{figure}

\begin{figure}
\begin{center}
\begin{subfigure}{0.45\textwidth}
  \includegraphics[width=6cm]{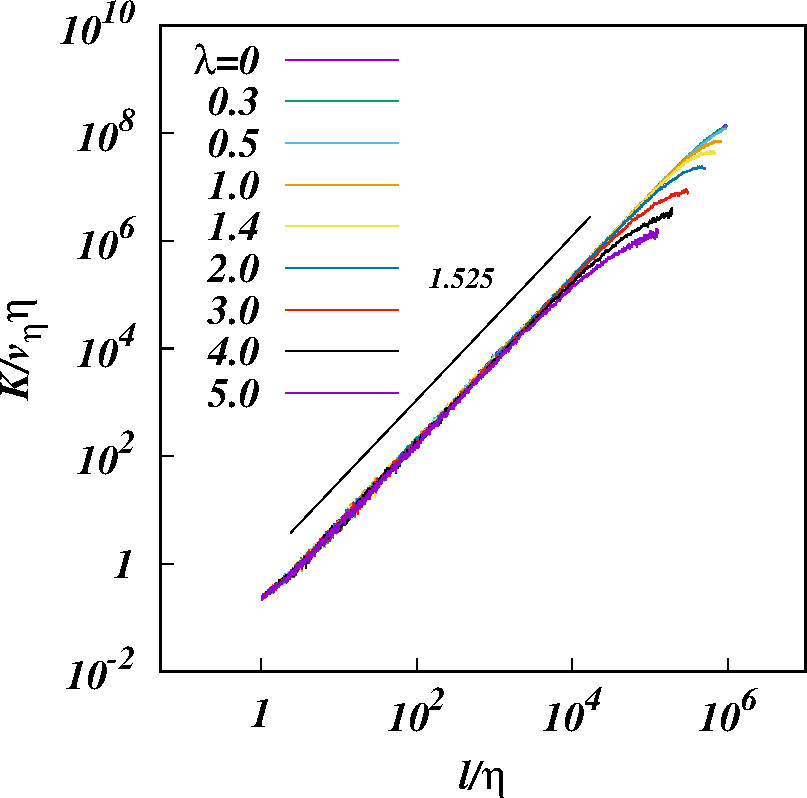}
   \caption{} 
\end{subfigure}
\begin{subfigure}{0.45\textwidth}
   \includegraphics[width=6cm]{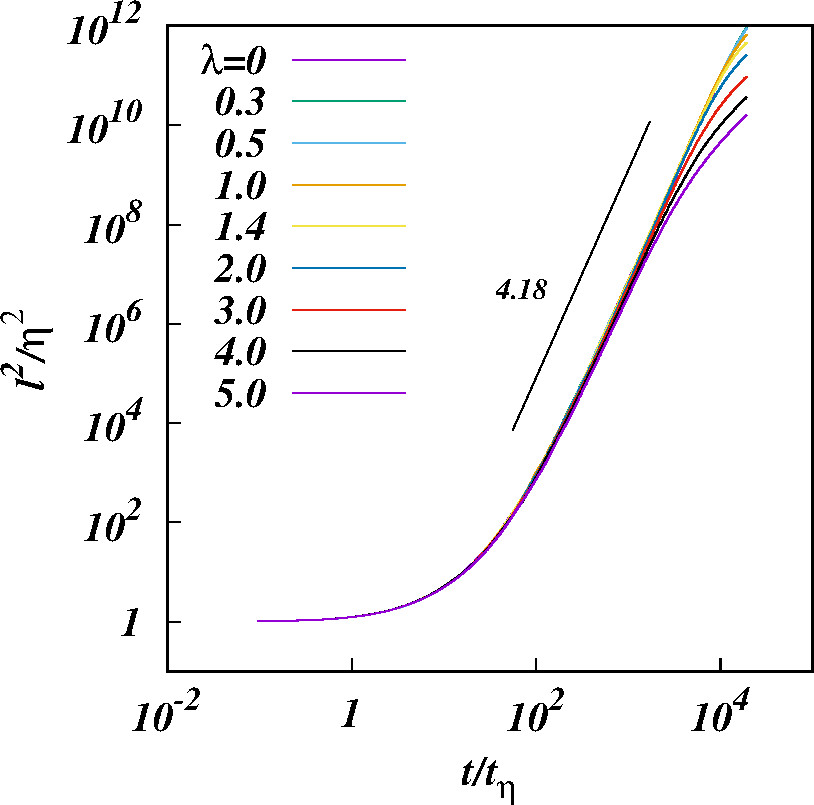}
  \caption{}
\end{subfigure}
\caption{(Colour online.)  Similar to Fig. \ref{fig_11}, except for and $R_k={10}^6$. } \label{fig_13}
\end{center}
\end{figure}

\begin{figure}
\begin{center}
\begin{subfigure}{0.45\textwidth}
  \includegraphics[width=6cm]{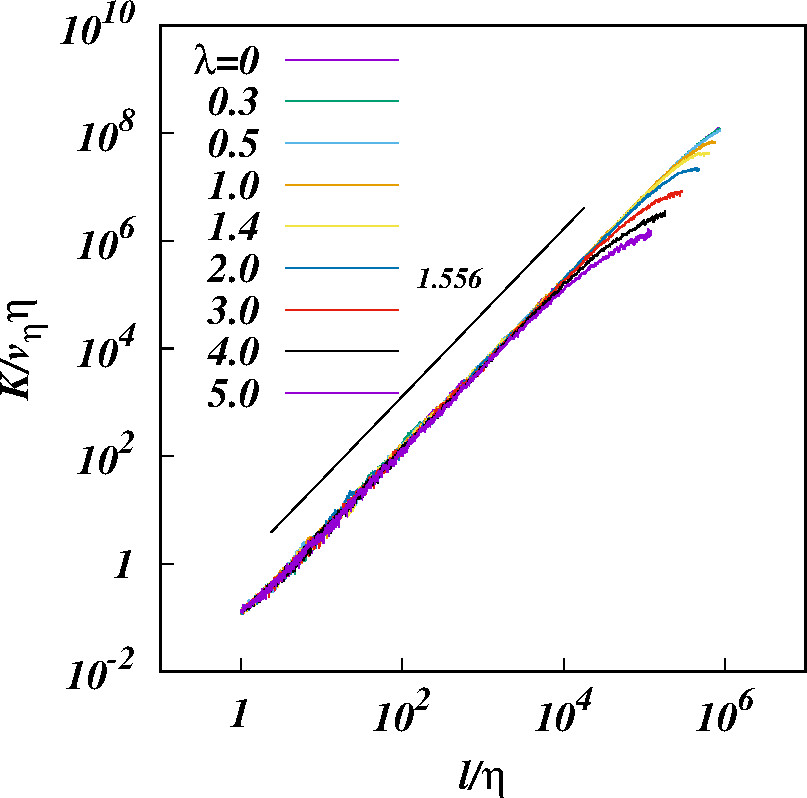}
   \caption{}  
\end{subfigure}
\begin{subfigure}{0.45\textwidth}
   \includegraphics[width=6cm]{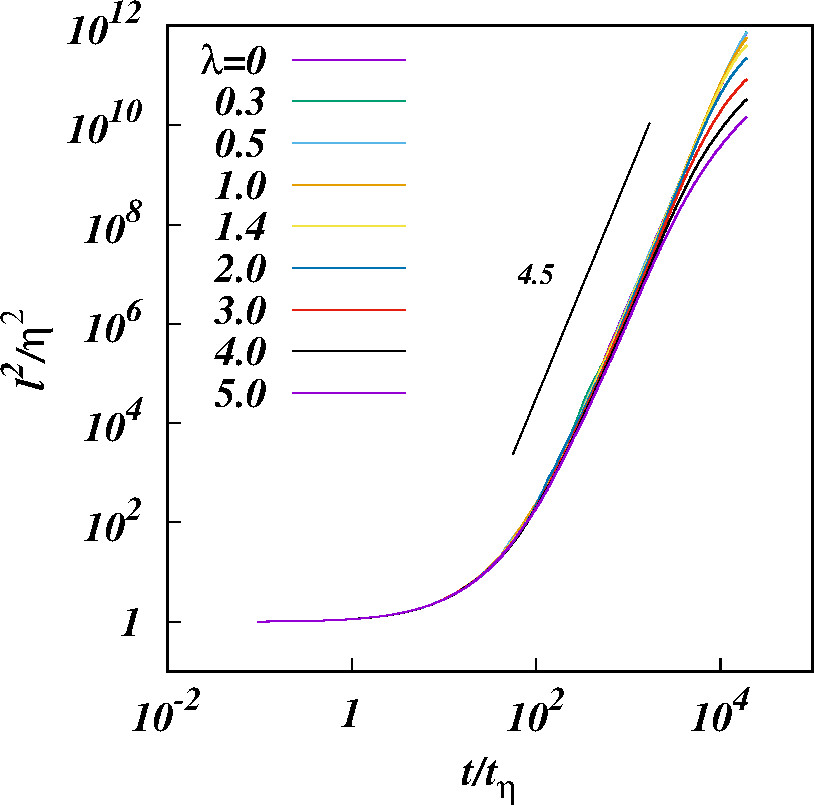}
  \caption{} 
\end{subfigure}
\caption{(Colour online.)  Similar to Fig. \ref{fig_11}, except for spectrum $E(k)\sim k^{-1.72}$ and $R_k={10}^6$. } \label{fig_14}
\end{center}
\end{figure}

{From equation (\ref{eq_4.2}),  we have
\begin{eqnarray}\label{eq_4.2b}
    \frac{K}{\eta v_\eta} &\sim&  \left({\frac{l}{\eta}}\right)^\gamma, \  0\ll l/\eta\ll R_k\to \infty.
\end{eqnarray}
From equation (\ref{eq_4.3}),  if we start with $l_0=\eta$ we obtain for small times,
\begin{eqnarray}\label{eq_4.4}
   \frac{ l^2}{\eta^2}  &\approx&  \frac{l_0^2}{\eta^2}=1, \qquad t/t_\eta \ll 1.
\end{eqnarray}
and for larger times, this is
\begin{eqnarray}\label{eq_4.5}
   \frac{ l^2}{\eta^2}  &\sim &   \left({\frac{t}{t_\eta}}\right)^{\chi},  \qquad  t/t_\eta\gg 1, \quad 1\ll l/\eta\ll R_k. 
\end{eqnarray}

At even larger times as the average separation increases outside of the inertial subrange, then each particle in a pair decorrelates from the other and the pair diffusion asymptotes towards two independent one-particle diffusion, \cite{Taylor1921},
\begin{eqnarray}\label{eq_4.6}
   { l^2}  &\to & 2   x^2(t) \sim 2(u')^2 T_L t,     \qquad t/t_\eta\gg 1, \quad l/\eta> R_k. 
\end{eqnarray}
where $u'$ is the rms turbulent fluctuation, and $T_L$ is the Lagrangian time scale. This is equivalent to a constant one-particle diffusivity since $K\sim 2d{\langle x^2\rangle}/dt\to $ constant  as $t/t_\eta\to \infty$. As $\lambda$ increases, the particles in a pair decorrelate at earlier times leading to earlier asymptote towards Taylor diffusion and a reduction in the separation scale when this occurs.

The different regimes are evident in the numerical results.  Fig. \ref{fig_11} shows the plots of,  (a)  $K/\eta v_\eta$ against $l/
\eta$, and (b) $ l^2/\eta^2$ against $t/t_\eta$, for $E(k)\sim k^{-5/3}$ and $R_k={10}^2$; several case of $\lambda$ are shown. Figs. \ref{fig_12} and \ref{fig_13} show the corresponding results for $R_k={10}^4$ and $R_k={10}^6$ respectively. Fig.  \ref{fig_14} shows the corresponding results for $E(k)\sim k^{-1.72}$,  $R_k={10}^6$.

In Fig. \ref{fig_11}, with $\gamma=1.41$, $\chi=2.94$, we observe  (a) $K/\eta v_\eta \sim (l/\eta)^{1.41}$, and in (b) $ l^2/\eta^2  \approx  1$ for small times, and $ l^2/\eta^2 \sim (t/t_\eta)^{2.94}$ at larger times. Similarly, 
in Fig.  \ref{fig_12}, (a) $K/\eta v_\eta \sim (l/\eta)^{1.5}$, and (b) $ l^2/\eta^2  \approx  1$ for small times, and $ l^2/\eta^2 \sim  (t/t_\eta)^{3.95}$ at larger times;
in Fig.  \ref{fig_13}, (a) $K/\eta v_\eta \sim (l/\eta)^{1.525}$, and (b) $ l^2/\eta^2  \approx  1$ for small times, and $ l^2/\eta^2 \sim  (t/t_\eta)^{4.18}$ at larger times;
in Fig.  \ref{fig_14}, (a) $K/\eta v_\eta \sim (l/\eta)^{1.556}$, and (b) $ l^2/\eta^2  \approx  1$ for small times, and $ l^2/\eta^2 \sim  (t/t_\eta)^{4.5}$ at larger times. See Tables 1 and 2 for the $\gamma$ and $\chi$ in the different cases.

In all the cases, the pair diffusion begins to approach Taylor diffusion at very long times. Note that  as the $\lambda$ increases the pair diffusion asymptotes towards Taylor diffusion at earlier times thus reducing the range of  separation over which the non-local regimes are valid and reducing the overall spread of particles, as predicted. 
}

{These results are very surprising  – in the small scales the diffusion is identical for frequencies in the  range $0\le\lambda\lesssim1$, and its only effect is to decrease the range of separations over which the scaling laws hold. Bigger sized turbulent motions containing most of the energy make bigger contributions to the pair relative motion at bigger separations – which was first noted by Richardson.  Increasing $\lambda$ thus reduces the particle spread only at  bigger separations. The results in Figs. \ref{fig_11}-\ref{fig_14} are consistent with this picture.}

The computed values for $\gamma$, and $\chi$, and the estimates for the constants,  $G_K$ and $G_l$, and their trends are also independent of $\lambda$ and therefore identical to the results in Tables \ref{Table2} and \ref{Table3}.

Based on these results we attempt to develop a plausible scenario that can rationalize these findings.
We know that (nearly all) streamlines in homogeneous and isotropic turbulence start and finish at infinity, thus allowing the spread of particles to infinity. (This is quite different from incompressible 2D turbulent streamlines which are closed.)
In 3D turbulent flows particles in pairs separate mainly in straining zones where the growth is exponential, but for a very short time. At other times, the particles move close together for long periods in parallel streaming zones, or they rotate close together in vortical zones – neither of these produces significant increase in the separation compared to the exponential growth in straining zones. 

Streamlines by themselves can diffuse particles, which could enhance both effects on particle separation growth rate: longer periods in  parallel streaming regions and in vortical regions could reduce the overall pair separation growth rate; on the other hand once the pair enters in to a straining zone their exponential growth rate also persists for a longer period of time, because the flow structures do not change, which could increase the overall pair separation. The net effect appears to be to leave the diffusion unchanged compared to the physically realistic cases where $0\le \lambda \lesssim 1$; this is possible if the underlying flow structures change slowly with respect to pair separation growth.

\begin{figure}
\begin{center}
\begin{subfigure}{0.9\textwidth}
   \centerline{\includegraphics[width=6cm]{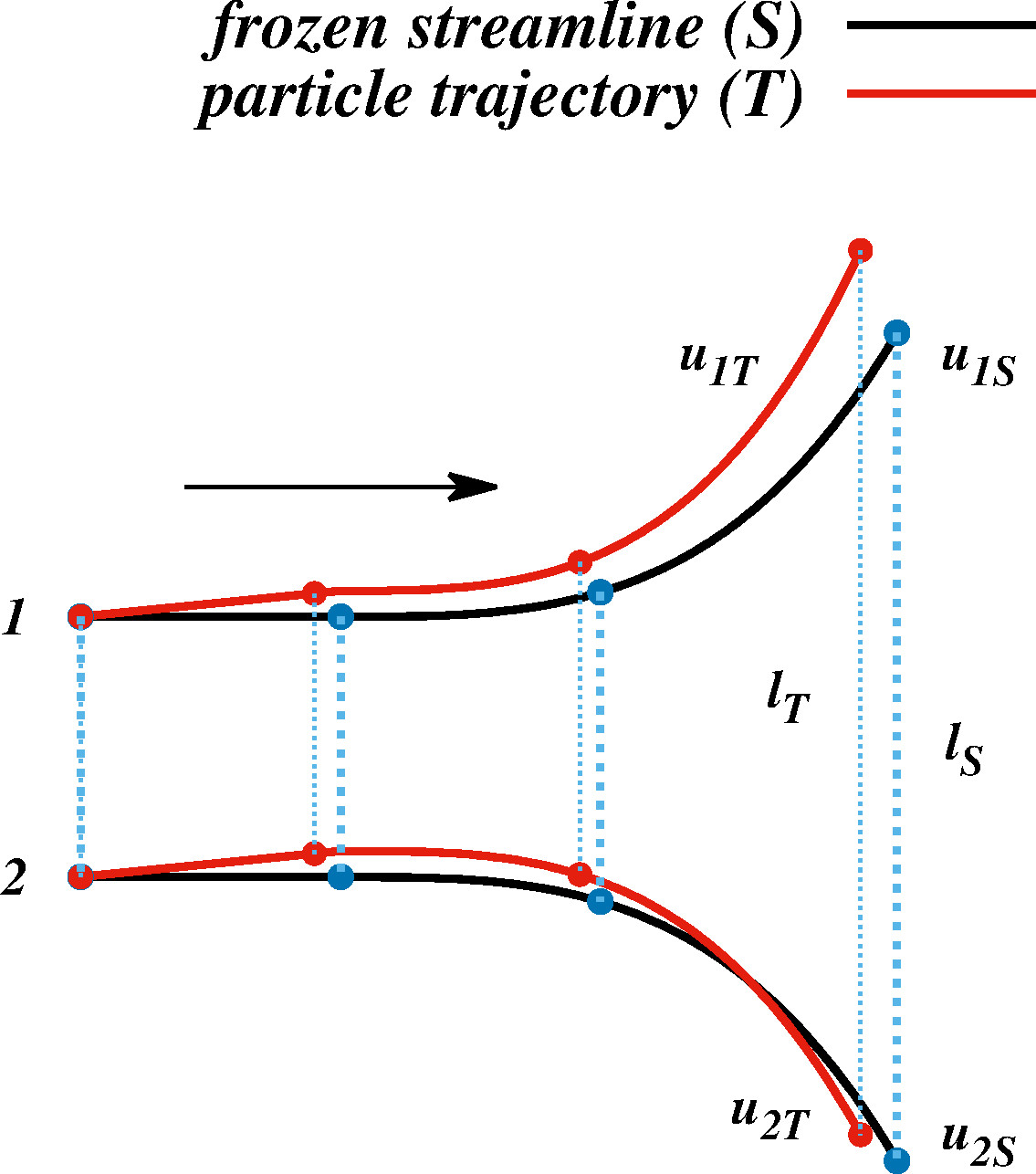}}
   \caption{Particle trajectories for, $0<\lambda\lesssim 1$  ($T$) , and for, $\lambda=0$  ($S$).}
\end{subfigure}\\
\vspace{5mm}
\begin{subfigure}{0.9\textwidth}
   \centerline{\includegraphics[width=6cm]{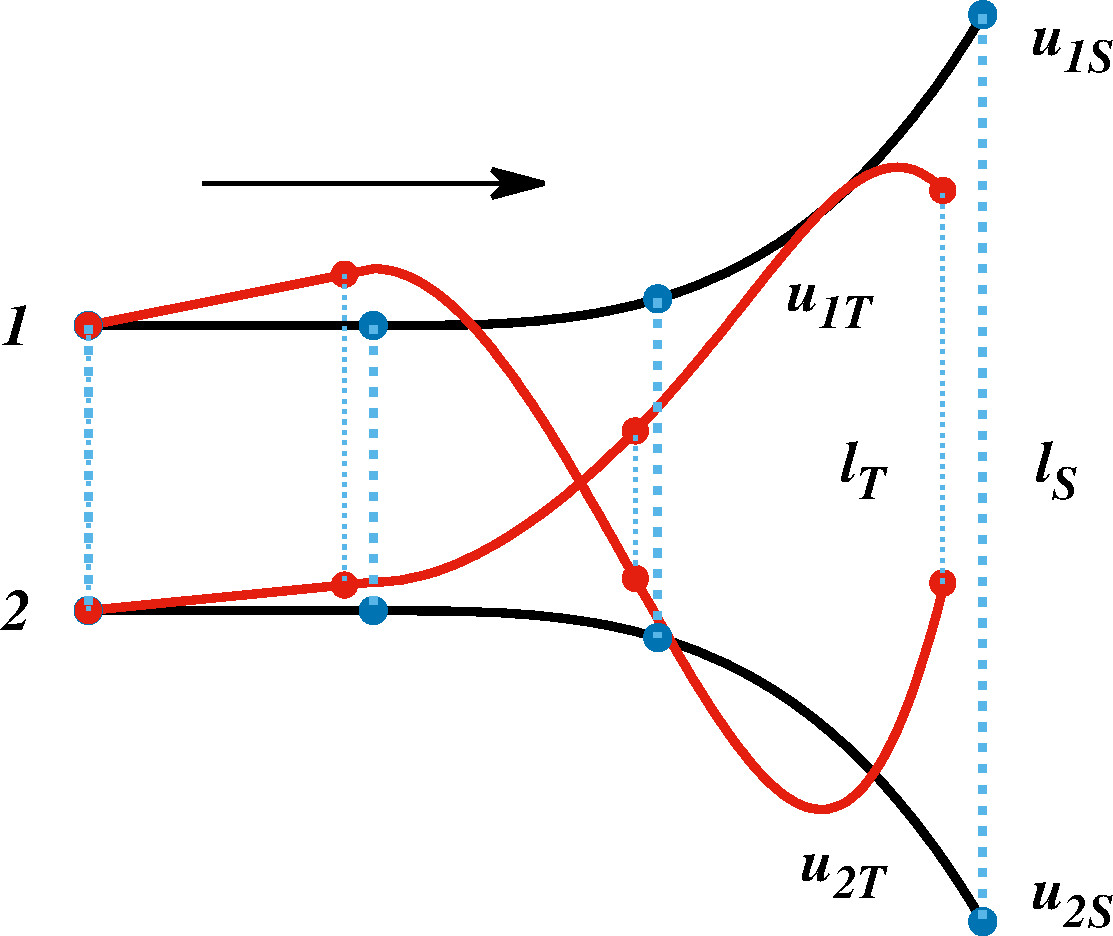}}
   \caption{Particle trajectories for ,  $\lambda\gg 1$  ($T$), and for, $\lambda=0$  ($S$).}
\end{subfigure}
\caption{(Color online.) {Sketches of  particle pairs (circles) and pairs of streamlines (1 and 2)  in the flow with $\lambda=0$ (S, black lines), and particle trajectories (T, red lines) in flow fields with:  (a)  $0<\lambda\lesssim 1$; (b) $\lambda\gg1$. The pair separations $l_S$ and  $l_T$ (dashed lines)  hardly differ in (a)  in 3D flow fields. The arrow indicates the general direction of particle motion.}}  \label{fig_15}
\end{center}
\end{figure}

Another viewpoint is that the ensemble of particle trajectories in 3D turbulent flows is probably very similar to the ensemble of streamlines, illustrated in Fig.  \ref{fig_15} which shows sketches of pairs of streamlines and trajectories for (a) $\lambda=0$ and $\lambda\lesssim 1$, and (b) $\lambda=0$ and $\lambda\gg 1$. Pairs of particles are continuously transported from one instantaneous pair of streamlines on to a nearby pair of streamlines. For $\lambda\lesssim 1$ in Fig.  \ref{fig_15}(a), the underlying flow structures change slowly, and hence the trajectories deviate slowly away from the initial streamlines, slowly enough that they  can be approximated by the streamlines themselves.  
Relative physical quantities, such as the pair relative velocity, ${\bf v}_T(t)={\bf u}_{2T}(t)-{\bf u}_{1T}(t)$, along pairs of particle trajectories differ very little from their counterparts along streamlines, ${\bf v}_S(t) = {\bf u}_{2S}(t)- {\bf u}_{1S}(t)$, and therefore the difference $|{\bf v}_T(t)- {\bf v}_S(t)|$ is much smaller than the first order quantity $|{\bf v}_{S}(t)|$, i.e.
\begin{eqnarray}\label{eq_5.2a}
    |{\bf v}_T(t)-{\bf v}_S(t)| &\ll& |{\bf v}_{S}(t)|
\end{eqnarray}
where the subscripts $T$ refers to the quantities along trajectories (red), and $S$ refers to the same quantities along the streamlines (black). Both sides of equation (\ref{eq_5.2a}) are small, but the left-hand side, being a second order difference of relative velocities, is much smaller than the right-hand side which is a simple first order difference of velocities.

Similarly, the pair separation satisfies,
\begin{eqnarray}\label{eq_5.3a}
    |{\bf l}_T(t)-{\bf l}_S(t)| &\ll& |{\bf l}_{S}(t)|.
\end{eqnarray}
because the separation along a pair of trajectories (T, red) $l_T$ is nearly the same as along a pair of streamlines (S, black) $l_S$, and therefore $l_T\approx l_S$,  in Fig. \ref{fig_15}(a).

\begin{figure}
\begin{center}
\begin{subfigure}{0.45\textwidth}
  \includegraphics[width=5cm]{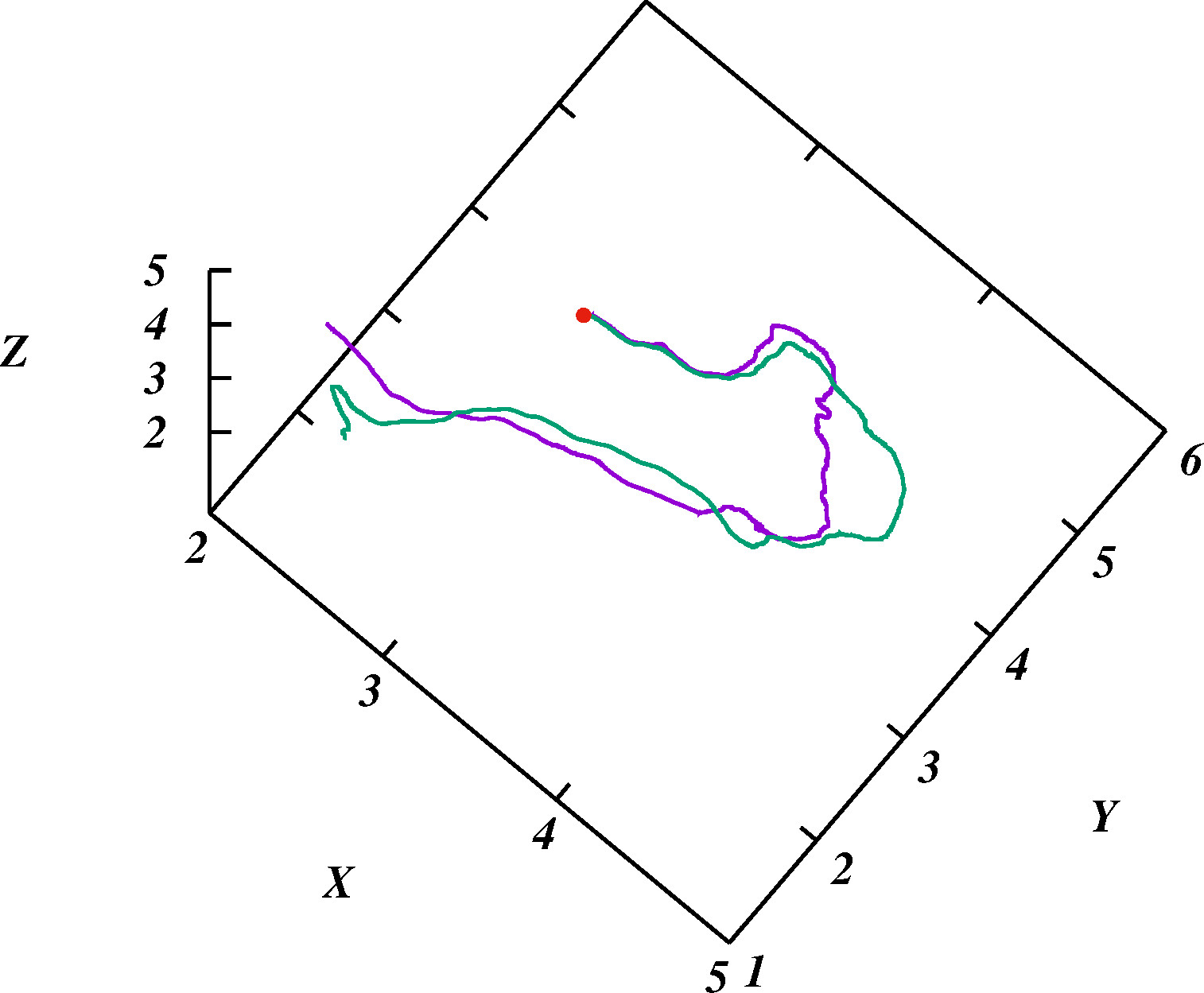}
   \caption{}    
\end{subfigure}
\begin{subfigure}{0.45\textwidth}
   \includegraphics[width=5cm]{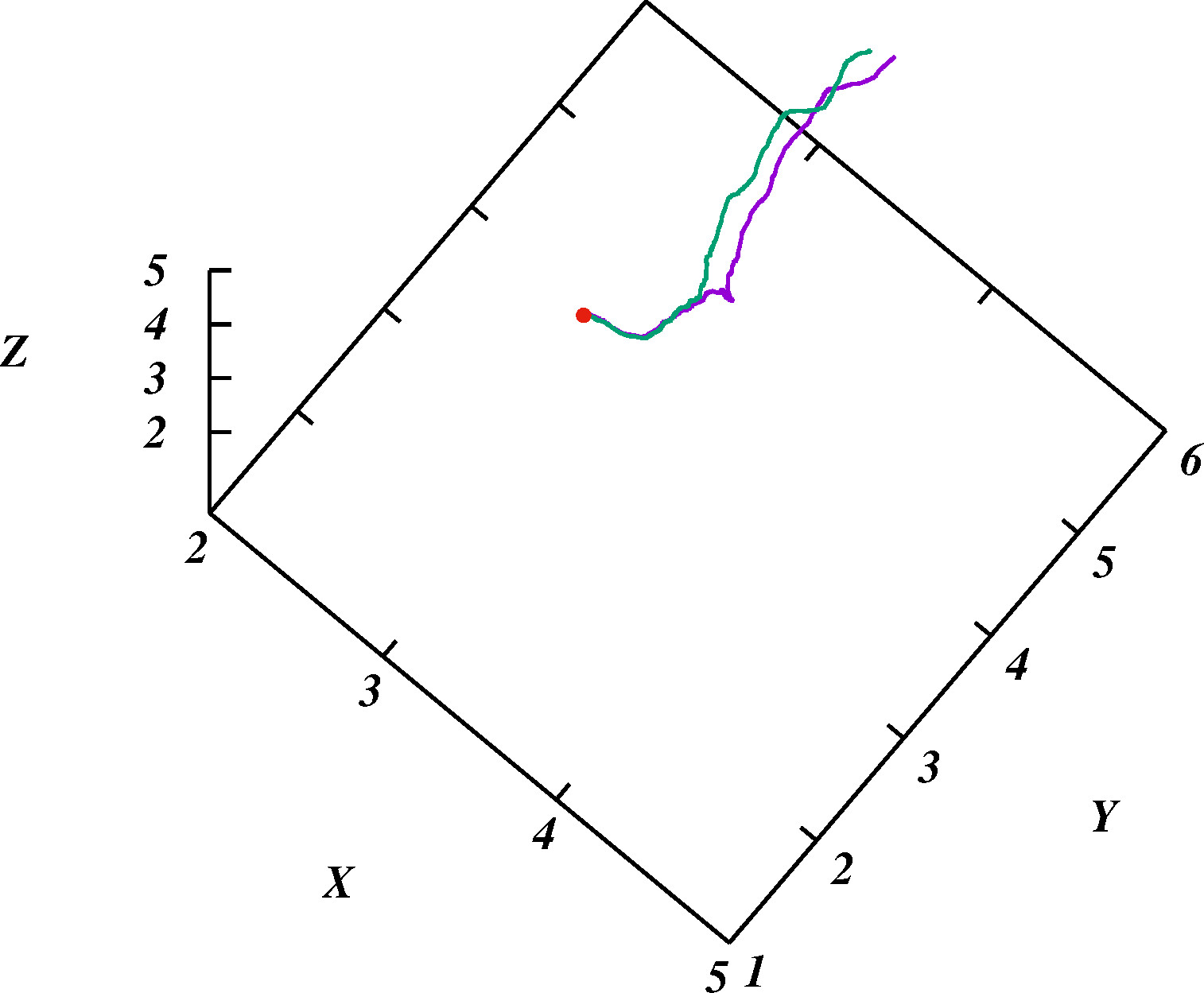}
  \caption{}
\end{subfigure}\\
\vspace{5mm}
\begin{subfigure}{0.45\textwidth}
  \includegraphics[width=5cm]{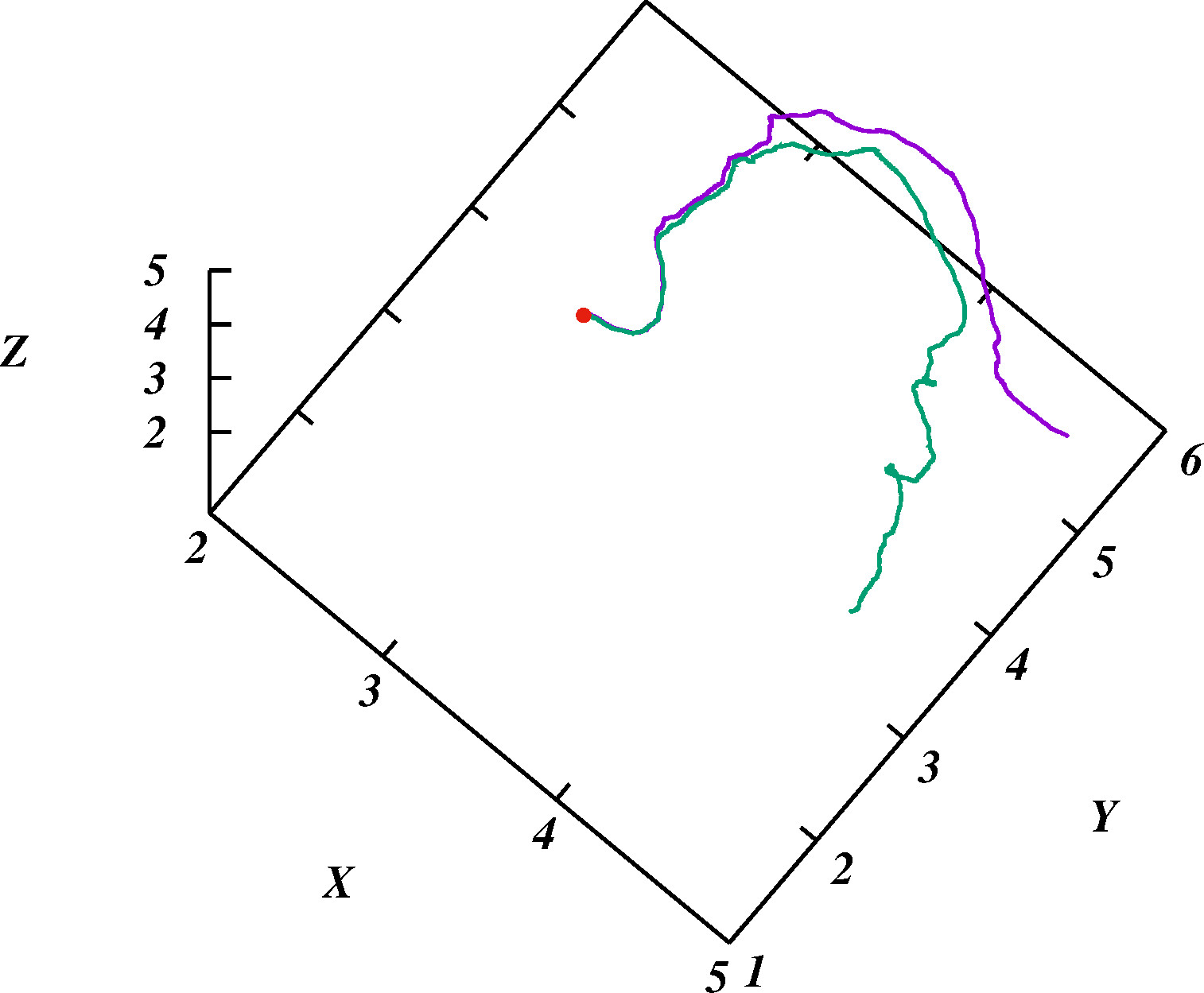}
   \caption{} 
\end{subfigure}
\begin{subfigure}{0.45\textwidth}
  \includegraphics[width=5cm]{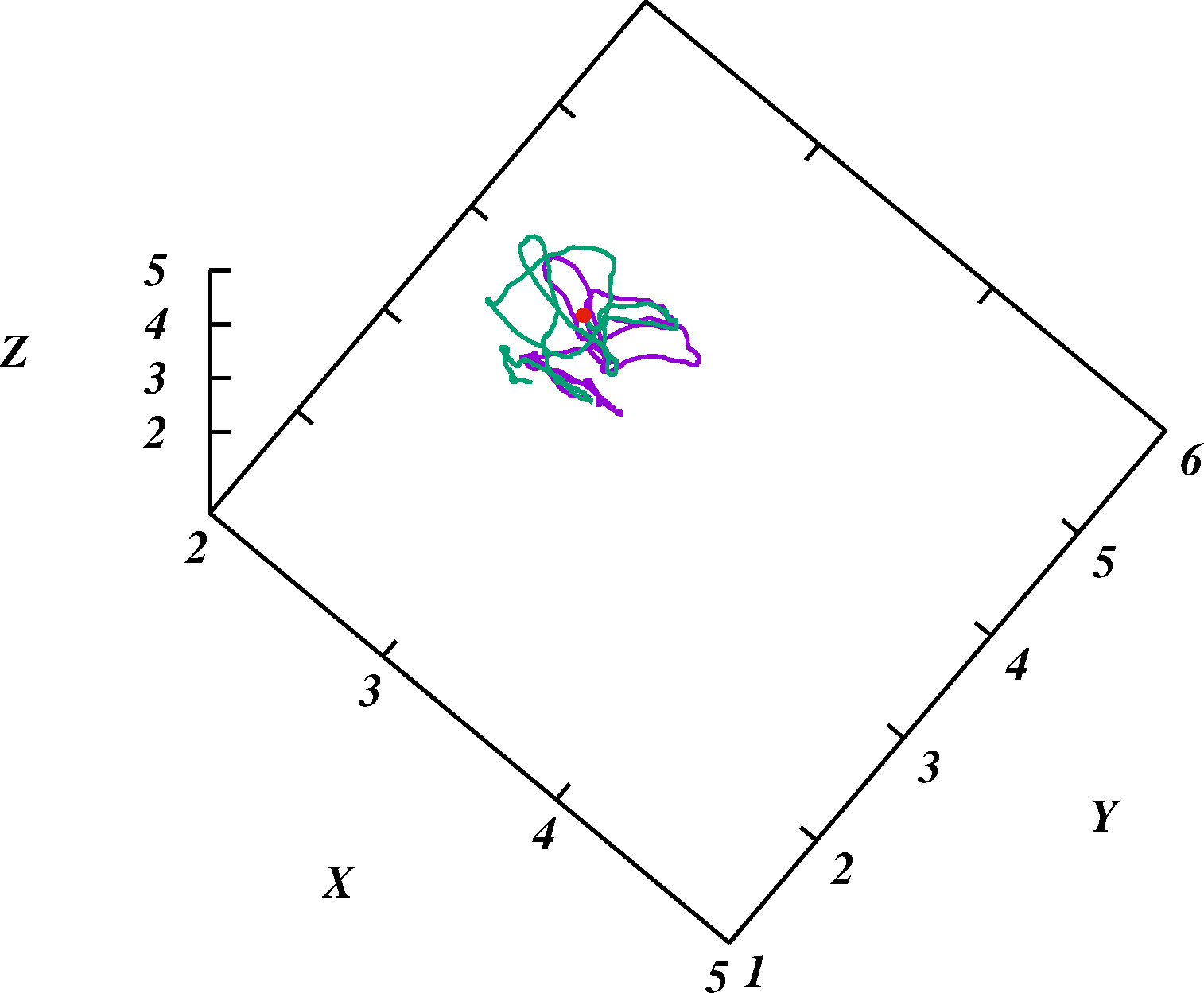}
   \caption{}  
\end{subfigure}
   \caption{{(Colour online.) Pairs of particle trajectories (purple and green) from the simulations with $E(k)\sim k^{-5/3}$ and $R_k=10^3$.  Raw units are shown, thus a distance of 6.2 is equivalent to $10^3\eta$. The different levels of $\lambda$ in each case is: (a) $0.0$, (b) $0.5$, (c) $1.0$, (d) $5.0$. The red circle marks the area where the particle pairs are released with separation $l_0=\eta$.}}    \label{fig_16}
\end{center}
\end{figure}

However, for $\lambda\gg 1$ in Fig.  \ref{fig_15}(b) this picture breaks down as the underlying flow structures change rapidly and the trajectories deviate significantly from the initial streamlines. Then, the pair relative velocity also changes rapidly approaching a random-walk like process, and we have $l_T\ll l_S$ (on average) thus decreasing the pair diffusion significantly.

\begin{figure}
\begin{center}
   \centerline{\includegraphics[width=5cm]{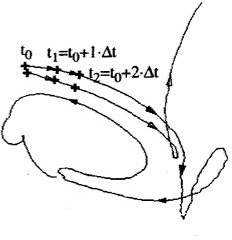}}
   \caption{Projections of particle trajectories in a plane, from Particle Tracking Velocimmetry experiments $Re = 11 600$, $U_{mean} = 148 mm/s$. $600$ timesteps. (From \cite{Virant1997}, with permission of Measurement Science and Technology).}    \label{fig_17}
\end{center}
\end{figure}

{This picture is evident in the pairs of particle trajectories generated in a single flow realization from KS, shown in Fig. \ref{fig_16}  for $E(k)\sim k^{-5/3}$ in an inertial subrange of $R_k=10^3$.  (Other cases display essentially the same features and are not shown.) Fig. \ref{fig_16} compares different cases of $\lambda=$ (a) 0.0, (b) 0.5, (c) 1.0, (d) 5.0. The pairs are released from the same point in space (red circle) with the same separation $l_0=\eta$. The long period of time during which the pairs  remain  close together before rapidly moving apart -- presumably on entering local regions of high strain --  are evident in Figs. \ref{fig_16}(a), (b), and (c).  The particle separations and the overall spread of the particles are of comparable magnitude in the three cases for $\lambda\le 1$, and it is notable how different these are from the case in Fig. \ref{fig_16}(d) for high $\lambda\gg 1$ where highly random particle relative motion leads to significantly reduced particle spread, as predicted above.

Further evidence in support of this process comes from the Particle Tracking Velocimetry (PTV) measurements in a channel by \cite{Virant1997} who used a 4-camera system and a 4-frame tracking algorithm based on the novel kinematic principle of minimum change in acceleration (\cite{Mass1993, Malik1993}). Fig. \ref{fig_17} shows the projection of particles moving in time in a turbulent channel flow in a plane. Note how the pair separation suddenly increases as the pair, presumably, enters a straining zone, and how similar this is to Figs. \ref{fig_16}(a), (b) and (c). The importance of Fig. \ref{fig_17} is heightened by the fact that,  to our knowledge, this is the only experimental visualization of  turbulent pair diffusion on the literature.
}

\section{Discussion and Conclusions}\label{Sec5}

The new physical picture fundamentally changes our understanding of the turbulent pair diffusion process and leads to new non-local scaling laws for pair diffusion. We stress that the non-local theory has support from measurements in both limits of small and large inertial subranges. In the limit $R_k\downarrow{10}^2$, the predicted quasi-local regimes $K\approx l^{4/3}$ have been observed with
$G_l\approx g_l\approx1$, which is consistent with most of the estimates in Table \ref{Table1} (Fig. \ref{fig_08}).
In the large inertial subrange limit, we predict $K\sim l^{1.556}$ as $R_k\to\infty$, which is within 1\% of the experimental scaling law from Richardson’s 1926 revised data, $K\sim l^{1.564}$, which lends considerable credibility to the non-local theory advanced here.

The celebrated `Richardson-Obukhov constant' $g_l$ is shown to be scale dependent and therefore not a constant as widely assumed, which explains the wide scatter in the estimated values of $g_l$ in the literature (it varies by two orders of magnitude). However, new constants associated with the diffusivity, $G_k$, and the pair separation (spread), $G_l$, are identified and they asymptote in our simulations to unexpectedly small values, $G_k\approx 0.73$ and $G_l\approx 0.01$.    

Another surprising finding from our work is that turbulent diffusion in the small scales is independent of the time scale of oscillation of the small scale flow structures.  We have rationalized that the this is because pair separation along pairs of particle trajectories remains almost unchanged compared to the separation along pairs of streamlines under realistic levels of flow field oscillations in time.

These results are significant for two reasons. First, they indicate the possible impact that the non-local theory could have in real world problems such as dispersal of droplets in a spray, pollution modelling, and the dispersal of viruses and bacteria and other biological contagions through airborne means --  less spread means higher concentrations sustained for longer times. 

Second,  the non-local theory indicates that local and non-local diffusional processes and very large-scale advection (sweeping) motions are approximately independent processes. Thus, a diffusion model based on a local-non-local-sweeping tri-variant process could be a fruitful approach to explore in the future. Furthermore, removing time dependence (of flow field oscillations) to model the statistics of turbulent diffusion in the small scales could greatly simplify and hence further improve modelling techniques.

{These findings also have important implications for the interpretation of pair diffusion studies. Some studies have reported, to a fair approximation,  $l^2 \sim t^3$. In the absence of a non-local theory this has been interpreted till now as a confirmation of R-O locality scaling laws. However, in our new picture $ l^2 \sim t^3$ is just the quasi-local limit for small inertial subranges. Actually, for real (i.e intermittent) turbulence it should be close to, $ l^2 \sim t^{3.13}$ from the purely local theory, or $ l^2 \sim t^{3.6}$ from the non-local theory which indicates that experiments and DNS still have a little way to go to achieving even quasi-local inertial subrange scaling. It is ironic that the quasi-local regime is predicted to appear in the limit $R_k\approx{10}^2$ where many current day measurements and DNS operate, presumably obscuring the greater significance of non-local diffusion. Further experiments and DNS are required to settle this matter.

The length scale $L_1$, the largest length scale in the inertial subrange, and $R_k$, the size of the inertial subrange, have played important roles in the scaling arguments and in characterizing the different pair diffusion regimes. Although we have argued that  $L_1$ could in principle be related to other more familiar length scales such as the integral length scale and the Taylor micro-scale, we believe that it is better to work in terms of $L_1$ and $R_k$ as fundamental parameters in turbulence theory and modelling. They have the merit of possessing clear physical meanings and they can be measured accurately in experiments and DNS.

Finally, we will have to wait for experiments and/or DNS to be realized in order to draw the last word on this issue. The results indicate that the transition from the quasi-locality limit at $R_k=10^2$ to the asymptotic non-local limit at $R_k=10^6$ is smooth. Thus at $R_k=10^3$ the predicted $K\sim l^{1.5}$ and $l^2\sim t^{3.7}$  scaling law in intermittent turbulence with $E(k)\sim k^{-1.72}$ (Table 2) should be observable -- we do not have to go all way to $R_k=10^6$ to test the main elements of the theory. Although $R_k=10^3$, may not seem too far from the current DNS of around $R_k=10^2$, in fact this is an increase in numerical grid size  of $10^3$, and with a timestep that is ten times smaller the simulation must run for ten times longer; this represents an increase in overall storage and costs of at least $10^4$ times bigger than today's DNS record achievement, and this will probably take quite some time to reach.\\
}

\noindent Declaration of Interests: the authors report no conflict of interest.

\section*{Acknowledgments}

\noindent The authors acknowledge the support of  the Information Technology Department at TTU for making available their  High Performance Computing facilities for this work.

\appendix
\section{Numerical Method}\label{AppA}

\subsection{Kinematic Simulations}\label{A1}

Kinematic Simulation (\cite{Kraichnan1970,Malik2017}) is a Lagrangian method for particle diffusion in which ensembles of random velocity filelds are produced as a sum of energy-weighted Fourier modes.  Lagrangian statistics are the physically meaningful output from KS.
There are two main physical inputs to KS. First, the Fourier coefficients are chosen such that they satisfy continuity exactly, hence mass is conserved by construction. Second, the square of the Fourier coefficients are proportional to the energy spectrum $E(k)$ which can be freely chosen within an arbitrary range of wavenumbers $k_1\le k\le k_\eta$. 
KS can thus generate extremely large inertial subrange.

KS has been used in turbulent diffusion studies for both passive and inertial particle motion, \cite{Maxey1987}, \cite{Turfus1987}, \cite{Farhan2015}. \cite{Meneguz2011} carried out a DNS of inertial particle motion, and compared it to results from KS which they found to  agree well with the DNS. \cite{Murray2016} investigated inertial particle statistics using KS. A recent application was in the investigation of the mid-latitude convective boundary layer (CBL) electricity above the homogeneous land surface using a large-eddy simulation (LES) where KS was used as a sub-grid scale model, see \cite{Anisimov2020}.

A KS flow field realization is produced as a truncated Fourier series,
\begin{eqnarray}\label{A1}
   {\bf U}({\bf x},t) =  \sum_{n=1}^{N}
   \left[{({\bf A_n\times \hat k_n})\cos{({\bf k_n\cdot x} +\omega_n t)} + 
          ({\bf B_n\times \hat k_n})\sin{({\bf k_n\cdot x} +\omega_n t)} }\right]
\ \ \ \
\end{eqnarray}
where $N$ is the number of representative wavemodes, typically hundreds for very long spectral ranges, $R_k=k_\eta/k_1\gg 1$. $\hat {\bf k}_n$ is a random unit vector (${\bf k}_n = \hat {\bf k}_n k_n$ and $k_n = |{\bf k}_n|$). The coefficients ${\bf A}_n$ and ${\bf B}_n$ are chosen such that their orientations are randomly distributed and uncorrelated with any other Fourier coefficient or wavenumber, and their amplitudes are
determined by $\langle {\bf A}^2_n \rangle = \langle {\bf B}^2_n \rangle \propto E(k_n)$, 
The angled brackets $\langle \cdot \rangle$ denotes the ensemble average over many flow fields. This ensures incompressibility in each flow realization, ${\bf \nabla} \cdot {\bf u} = 0$. The flow field ensemble generated in this manner is statistically homogeneous, isotropic, and  stationary.

{An important feature of KS is that unlike some other Lagrangian methods, by generating entire kinematic flow fields in which particles are tracked it does not suffer from the crossing-trajectories error which is caused when two fluid particles occupy the same location at the same time in violation of incompressibility; but because KS flow fields are incompressible by construction this error is completely eliminated.}

In turbulent particle pair studies the interest is in Kolmogorov-like power law spectra,
\begin{eqnarray}\label{A2}
      E(k) &=& C_E \varepsilon^{2/3}L^{5/3-p}k^{-p}, \ \ k_1\le k\le k_\eta, \ \ 
      1<p\le 3
\end{eqnarray}
$C_E$ is a constant. The largest represented scale of turbulence is $2\pi/k_1$, and the smallest is the Kolmogorov scale $\eta=2\pi/k_\eta$.
The constant is normalized such that the total energy contained in the range is $3(u')^2/2$, where $u'$ is the rms turbulent velocity fluctuations in each direction.  $\varepsilon(p)$ is determined by integrating the spectrum, $\int_{k_1}^{k_\eta} E(k)dk=3(u')^2/2$. ($p=1$ is a singular limit which is not consider here.)  For Kolmogorov turbulence, the time microscale $v_\eta = (\varepsilon\eta)^{1/3}$ is the velocity microscale, and $\tau_\eta = \varepsilon^{-1/3} \eta^{2/3}$. 

The distribution of the wavemodes is geometric, $k_n=k_1 r^{n-1}$, with $r=(k_\eta/k_1)^{1/(N-1)}$. The grid size in wavemode-space of  the $n^{th}$ wavemode is $\delta k_n = k_n(\sqrt{r}- 1/\sqrt{r})$.

The usual practice is to make the $n$'th mode frequency proportional to the eddy-turnover frequencies,  
\begin{eqnarray}\label{A3}
    \omega_n &=& \lambda\sqrt{k_n^3E(k_n)}. 
\end{eqnarray}
The choice of $\lambda$ is arbitrary, although $\lambda = 0.5$ is a common choice.

A particle trajectory is obtained by integrating the Lagrangian velocity ${\bf U}_L(t)$,
\begin{eqnarray}\label{A4}
   {d{\bf x} \over dt} = {\bf U}_L(t) = {\bf U}({\bf x},t),
\end{eqnarray}
{using any standard method such as Runga-Kutta or Predictor-corrector methods. The  time step $\Delta t$ should be smaller than any other time scale in the system -- in this case the Kolmogorov time scales; thus we require $\Delta t \ll \tau_k$. In our simulations, particle trajectories are produced by integrating equation (\ref{A4}) with a fixed time step of $\Delta t \approx 0.01\tau_k$ in a fourth order Adams-Bashforth predictor-corrector method.   

It is important to produce a very large ensemble of independent particle pair trajectories. Eight independent particle pairs are released in each flow realization, each pair set part by more than an integral length scale. Each pair is initially released with a pair separation distance of $l_0/\eta\ge 1$. To obtain a true ensemble, this process is repeated in many thousands of KS flow realizations. (Releasing thousands of particle pairs in the same KS flow field will not produce the required independence which may produce biased statistics.) Pairs of trajectories are thus harvested over a large ensemble of flow realizations and pair statistics are then obtained from it for analysis.

KS has been validated against DNS in \cite{Malik1999} where pair diffusion  statistics up to 4th order (flatness) were found to be close to the DNS data.
}

\subsection{Thomson \& Devenish's conjecture}\label{A2}

KS has not been without its critics. \cite{Thomson2005,Nicolleau2011,Eyink2013} argued that the convection of the small scales of turbulence by larger scales of motions is important in turbulent diffusion and KS does not contain such `sweeping' action. They noted that KS gave scaling laws for the pair diffusivity that deviates from locality and they concluded that this was evidence that KS is therefore inaccurate for turbulent diffusion studies. See also \cite{Nicolleau2011,Eyink2013}.

However, the authors' arguments cannot be sustained under detailed examination and has been disproved in \cite{Malik2017} with mathematical rigor. Here we recount the important steps in the counter argument. \cite{Thomson2005} claim to have obtained KS scaling laws theoretically that match their KS results. This appears to be serendipitous which unfortunately obscures the broader picture. (Actually, the scaling law that they predict, $\sim l^{1.555}$, does not quite match the KS scaling $\sim l^{1.525}$ in equation (\ref{eq_2.8}) for Kolmogorov turbulence which is the only case that they considered, but it is close to the case with intermittency in equation (\ref{eq_2.10})). Their derivation is based on the dimensionality of the relative pair diffusivity, $[K]=U^2T$ where $U$ is velocity and $T$ is time, from which they assume that $U^2$ scales like the Eulerian structure function $S(l)=\langle\Delta u^2\rangle \sim \varepsilon^{2/3} l^{2/3}$, and that $T$ scales with the local timescale $T\sim \tau(l)$; thus 
\begin{eqnarray}\label{A5}
   K(l) \sim S(l)\tau(l) \sim \varepsilon^{2/3} l^{2/3} \tau(l).
\end{eqnarray}
They argue that the timescale $\tau(l)$ in KS is not Kolmogorov $\tau(l)\sim l^{2/3}\varepsilon^{1/3}$, and this leads to `unphysical' scaling laws. However, equation (\ref{A5}) remains an unproven conjecture.

We stress that the arguments made by the authors were based on the firm conviction that locality {\em must} be true in turbulent pair diffusion -- but as we have seen this is a flawed assumption. Furthermore, although it is true that KS does not contain dynamical sweeping action, the more important point is to quantify this error. \cite{Thomson2005} consider two case, one of which contains a constant mean flow $U>0$, and in the second case the mean flow is zero, $U=0$; we consider only the latter case here. In \cite{Malik2017} a detailed mathematical analysis was carried out and the sweeping errors were quantified  for the first time through the relative error $e_K(l)=\Delta K/K$ where $\Delta K$ is the error in the pair diffusivity due to the lack of sweeping, divided by the pair diffusivity itself; it was found that it is negligible $|e_K|\ll 1$ inside the inertial subrange, even for very large inertial subrange, and the deviations from locality laws  seen in KS  are therefore physically genuine.

\section{Existence of the non-local neigbourhood}\label{AppB}

\begin{figure}
\begin{center}
\begin{subfigure}{0.45\textwidth}
  \includegraphics[height=5cm]{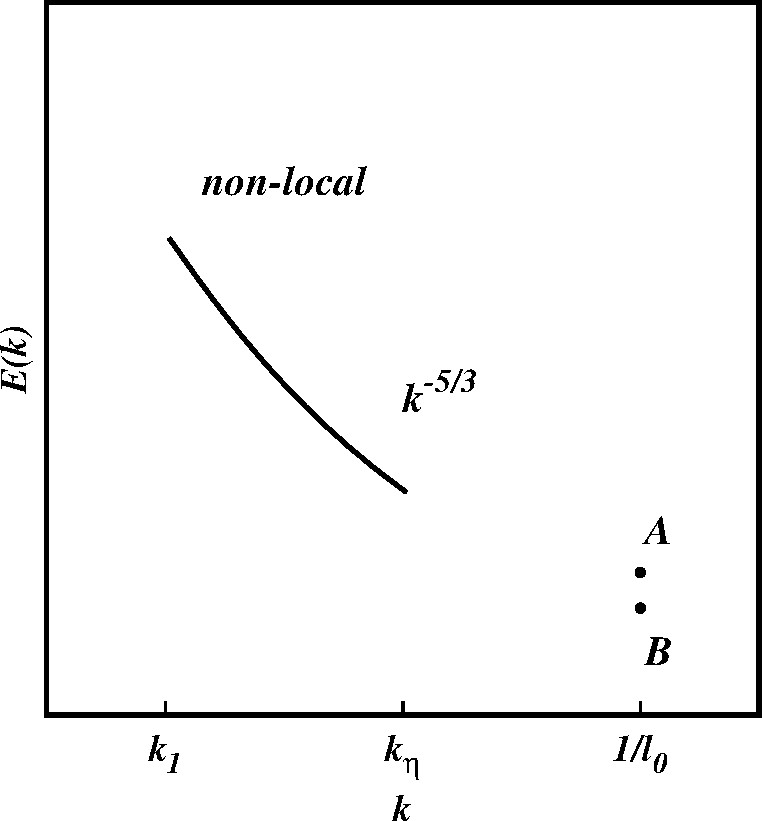}
   \caption{}   
\end{subfigure}
\begin{subfigure}{0.45\textwidth}
   \includegraphics[height=5cm]{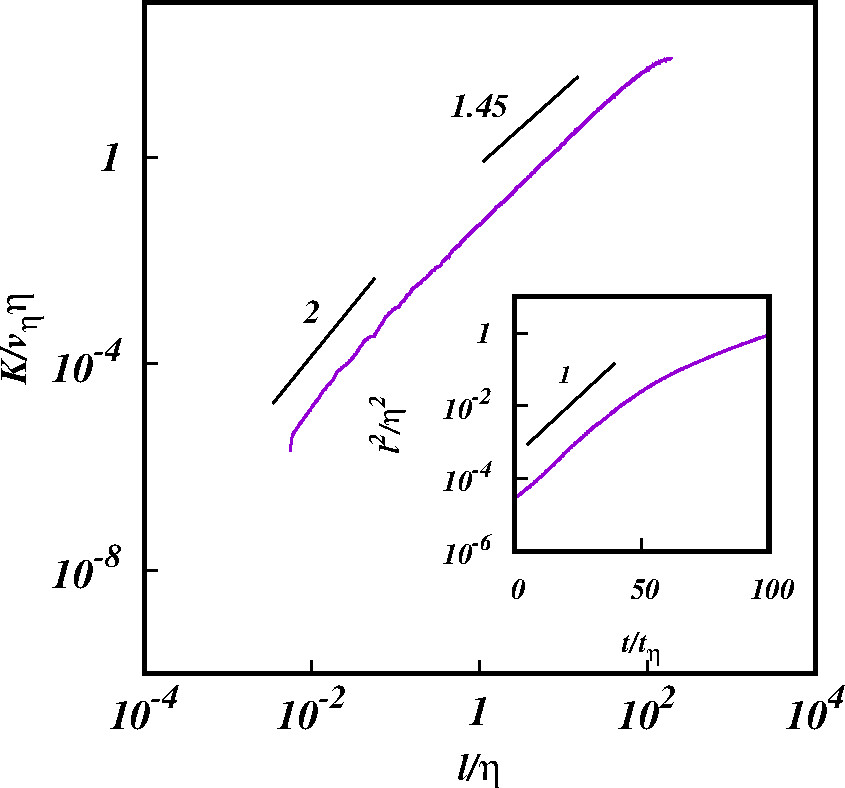}
  \caption{} 
\end{subfigure}\\
   \caption{(Colour online). {(a) Sketch of the Kolmogorov spectrum used in this KS with a spectral gap between $k_\eta\le k\le 1/l_0$ so that initially the entire spectrum is `non-local' to the particle pair A and B with separation  $l_0\ll\eta$. (b) log-log plot of $K/\eta v_\eta$ against $l/\eta$; inset, log-linear plot of $l^2/\eta^2$ against $t/t_\eta$ for early time showing exponential growth before entering inside the inertial subrange. Lines of different slopes as indicated for comparison.}}    \label{fig_18}
\end{center}
\end{figure}

The existence of a non-local neighbourhood can be demonstrated explicitly. We release an ensemble of particle pairs with initial separation much smaller than the Kolmogorov scale, $l_0\ll \eta$. There is then a large spectral gap between the initial pair separation and the Kolmogorov scale, $k_\eta\le k \le 1/l_0$ where we would normally expect the `local' range of  wavenumber. Only the non-local range of  wavenumbers (now $k_1\le k\le k_\eta$) exists initially, as illustrated in Fig. \ref{fig_18}(a). The pair separation will grow under the influence of the non-local processes alone, at least until the average separation approaches $L_1$, $l\uparrow L_1$.  For convenience we use Kinematic Simulations (KS)  (see Appendix A) data to delineate our theory and compare results for different sizes of the inertial subrange. Results with  $R_k=k_\eta/k_1=300$, and $l_0=10^{-2}\eta$ are shown in Fig. \ref{fig_18}(b) for the pair diffusivity $K$ against $l$ where we observe $K\sim l^2$ at early times when $l\ll\eta$ which is characteristic of exponential growth, before the curve shifts at later times towards a shallower power law, characteristic for this inertial subrange (Section 4, Table \ref{Table2}). The inset in this figure is the log-linear plot of the pair  separation $l^2$ against time $t$ at early times, again clearly showing exponential growth (the straight line). All other cases, for different $R_k$ and $E(k)\sim k^{-p}$, display similar trends and are not shown. (In actual turbulent flow we expect viscous diffusion will slightly decrease the separation growth rate.)

\begin{thebibliography}{14}
\expandafter\ifx\csname natexlab\endcsname\relax\def\natexlab#1{#1}\fi

\bibitem[Anisimov et al(2020)]{Anisimov2020}
{\sc Anisimov S. V., Galichenko S. V., Prokhorchuk A. A. \& Aphinogenov} 
Mid-latitude convective boundary-layer electricity: A study by large-eddy simulation.
{\em Atmospheric Research}, 244:105035 (2020).

\bibitem[Anselment et al(2001)]{Anselment2001}
{\sc Anselment F., Antonia R. A., Danaila L.} 
Turbulent flows and intermittency in laboratory experiments. 
{\em Planetary and Space Science}, 49:1177–1191 (2001).

\bibitem[Bec et al(2010)]{Bec2010}
{\sc Bec, J., Biferale, L., Lanotte, A., Scagliarini, A. \& Toschi, F.}
Turbulent pair dispersion of inertial particles.
{\em J. Fluid Mech.} 645, 497-528, (2010).

\bibitem[Batchelor(1953)]{Batchelor1953}
{\sc Batchelor, G. K}
The theory of homogeneous Turbulence.
{\em Cambridge University Press} (1953).

\bibitem[Berg et al(2006)]{Berg2006}
{\sc Berg J., Luthi B., Mann J., Ott S.}
Backwards and forwards relative dispersion in turbulent flow: an experimental investigation. 
{\em Phys. Rev. E} 74:016304, (2006).

\bibitem[Biferale et al(2005)]{Biferale2005}
{\sc Biferale L., Boffetta G., Celani A., Devenish B. J., Lanotte A., Toschi F.}
Lagrangian statistics of particle pairs in homogeneous isotropic turbulence. 
{\em Phys. Fluids} 17:115101, (2005).

\bibitem[Boffetta \& Sokolov(2002)]{Boffetta2002}
{\sc Boffetta G. \& Sokolov I. M.}
Relative dispersion in fully developed turbulence: the Richardson’s law and intermittency corrections. 
{\em Phys. Rev. Lett.} 88:094501, (2002).

\bibitem[Borgas \& Sawford(1994)]{Borgas1994}
{\sc Borgas M. S. \& Sawford B. L.}  
A family of stochastic models for two-particle dispersion in isotropic homogeneous stationary turbulence.
{\em J. Fluid Mech.} 279:69–99, (1994).

\bibitem[Borgas \& Yeung(1998)]{Borgas1998}
{\sc Borgas M. S. \& Yeung P. K.}
Conditional fluid-particle accelerations in turbulence. 
{\em Theor. Comput. Fluid Dyn.} 11:69–93, (1998).

\bibitem[Buaria et al(2015)]{Buaria2015}
{\sc D. Buaria, Brian L. Sawford, and P. K. Yeung}
Characteristics of backward and forward two-particle relative dispersion in turbulence at different Reynolds numbers
{\em Phys. Fluids,\/} 27:105101 (2015).

\bibitem[Durbin(1980)]{Durbin1980}
{\sc Durbin, P.} 
A stochastic model of two-particle dispersion and concentration fluctuations in homogeneous turbulence. 
{\em J. Fluid Mech.,\/} {100(2)}, 279-302. (1980).

\bibitem[Darragh et al(2020)]{Darragh2020}
{\sc Ryan Darragh, Colin A. Z. Towery, Alexei Y. Poludnenko \& Peter E. Hamlington} 
Particle Pair Dispersion and Eddy Diffusivity in a High-Speed Premixed Flame. 
{\em Phys. Fluids,\/} {Submitted.} (2020).

\bibitem[Elliott \& Majda(1996)]{Elliott1996}
{\sc Elliott F. W. \& Majda A. J.} 
Pair dispersion over an inertial range spanning many decades. 
{\em Phys. Fluids} 8:1052–60, (1996).

\bibitem[Eyink \& Benveniste(2013)]{Eyink2013}
{\sc Eyink, G. L. \& Benveniste, D.} 
Suppression of particle dispersion by sweeping effects in synthetic turbulence.
{\em Phys. Rev. E\/} {87}, 023011 (2013).

\bibitem[Farhan et al(2015)]{Farhan2015}
{Farhan M, Nicolleau FCGA, Nowakowski AF.}
Effect of gravity on clustering patterns and inertial particle attractors in kinematic simulations.  
{\em Phys. Rev. E\/} 91:043021 (2001).

\bibitem[Franzese \& Cassiani(2007)]{Franzese2007}
{\sc Franzese P. \& Cassiani M.}
A statistical theory of turbulent relative dispersion. 
{\em J. Fluid Mech.} 571:391–417, (2007)

\bibitem[Hentschel et al(1984)]{Hentschel1984}
{Hentschel H.  G. E. \& Procaccia I.}
Relative diffusion in turbulent media: the fractal dimension of clouds.
{\em Phys. Rev. A\/} {29}, 1461–1470. (1984).

\bibitem[Heppe(1998)]{Heppe1998}
{\sc Heppe B. M. O.}
Generalized Langevin equation for relative turbulent dispersion.
{\em J. Fluid Mech.} 357:167–98, (1998).

\bibitem[Hussain \& Stout(2013)]{Hussain2013}
{\sc Hussain F. \& Stout E.}
Generalized Langevin equation for relative tSelf-limiting and regenerative dynamics of 
perturbation growth on a vortex columnurbulent dispersion.
{\em J. Fluid Mech.} 718:39–88, (2013).

\bibitem[Hunt(1985)]{Hunt1985}
{\sc Hunt J. C. R.}
Turbulent Diffusion from Sources in Complex Flows.
{\em Ann . Rev. Fluid Mech.} 17:447–485, (1985).

\bibitem[Ishihara \&  Kaneda(2002)]{Ishihara2002}
{\sc Ishihara T. \&  Kaneda Y.}
Relative diffusion of a pair of fluid particles in the inertial subrange of turbulence.
{\em Phys. Fluids} 14:L69–72, (2002).

\bibitem[Julian et al(1977)]{Julian1977}
{\sc Julian P., Massman W., Levanon N.}
The TWERL experiment. 
{\em Bull. Am. Meteorol. Soc.} 58:936–48, (1977).

\bibitem[Klafter et al(1987)]{Klafter1987}
{Klafter J, Blumen A, Shlesinger MF.}
Stochastic pathway to anomalous diffusion.  
{\em Phys. Rev. A\/} {35}, 3081–85.  (1987).

\bibitem[Kraichnan(1966)]{Kraichnan1966}
{\sc Kraichnan R. H.}
Dispersion of particle pairs in homogeneous turbulence. 
{\em Phys. Fluids} 9:1937–43, (1966).

\bibitem[Kraichnan(1970)]{Kraichnan1970}
{\sc Kraichnan R. H.}
Diffusion by a random velocity field. 
{\em Phys. Fluids} 13, 22-31 (1970). 

\bibitem[Kurbanmuradov \& Sabelfeld(1995)]{Kurbanmuradov1995}
{\sc Kurbanmuradov O. A. \& Sabelfeld K. K.}
Stochastic Lagrangian models of relative dispersion
of a pair of fluid particles in turbulent flows. 
{\em Monte Carlo Methods Appl.} 1:101–36, (1995).

\bibitem[Larcheveque \& Lesieur(1981)]{Larcheveque1981}
{\sc Larcheveque M. \& Lesieur M.}
The application of eddy-damped Markovian closures to the problem of dispersion of particle pairs. 
{\em J. M'ec.} 20:113–34, (1981).

\bibitem[Lesieur(1981)]{Lesieur1981}
{\sc Lesieur M.} 
Dispersion of particle pairs in homogeneous turbulence. 
{\em Phys. Fluids} 9:1937–43, (1981).

\bibitem[Lundgren(1981)]{Lundgren1981}
{\sc Lundgren T. S.} 
Turbulent pair dispersion and scalar diffusion. 
{\em J. Fluid Mech.} 111:27–57, (1981).

\bibitem[Ma et al(2020)]{Ma2020}
{\sc Ma D., Jianmin G., Zhang Z., Zhaoa H. \& Wang Q.} 
Locating the gas leakage source in the atmosphere using the dispersion wave method.
{\em J. Loss Prevent. Proc. Indust.\/} {12(12)}:e0189917, (2017).

\bibitem[Malik \& Vassilicos(1999)]{Malik1999}
{Malik N. A. \& Vassilicos J. C.}
A Lagrangian model for turbulent dispersion with turbulent-like flow structure: comparison with direct numerical simulation for two-particle statistics.
{\em Phys. Fluids\/} 11:1572-1580 (1999). doi: 10.1063/1.870019.

\bibitem[Malik(2017)]{Malik2017}
{\sc Malik, N. A.} 
Residual sweeping errors in turbulent particle pair diffusion in a Lagrangian diffusion model.
{\em PLoS ONE\/} {12(12)}:e0189917, (2017).

\bibitem[Malik et al(1993)]{Malik1993}
{\sc Malik N. A., Dracos Th. \& Papantoniou, D.} 
{Particle Tracking Velocimetry in Three Dimensional Turbulent Flows. Part II: Particle Tracking.
{\em Experiments in Fluids} 15(4), 279—294, (1993).

\bibitem[Mann et al(1999)]{Mann1999}
{\sc Mann J., Ott S. \& Andersen J. S.}
Experimental study of relative turbulent diffusion.
{\em Risø-R-1036(EN)}, Risø Nat. Lab., Roskilde, Denmark, (1999).

\bibitem[Mass et al(1993)]{Mass1993}
{\sc Mass, H. G., Gruen, A. \& Papanatonio, D.}
Particle Tracking Velocimetry in Three Dimensional Turbulent Flows. Part 1. Photogrammetric determination of particle coordinates.
{\em Experiments in Fluids} 15(2), 133—146, (1993).

\bibitem[Maxey(1987)]{Maxey1987}
{Maxey M. R.} 
The gravitational settling of aerosol particles in homogeneous turbulence and random flow fields.
{\em J. Fluid Mech.\/} 174:441-465 (1987).

\bibitem[Meyers \& Meneveau(2008)]{Meyers2008}
{\sc Meyers J., \& Meneveau C.} 
A. Functional form for the energy spectrum parametrizing bottleneck and intermittency effects. 
{\em Phys. Fluids}, 20:065109 (2008).

\bibitem[Meneguz \& Reeks(2011) ]{Meneguz2011}
{Meneguz E. \& Reeks M. W.}
Statistical properties of particle segregation in homogeneous isotropic turbulence.
{\em J. Fluid Mech.\/} 686:338-351 (2011).

\bibitem[Murray et al(2016)]{Murray2016}
{Murray S., Lightstone M. F. \& Tullis S.}
Single-particle Lagrangian and structure statistics in kinematically simulated particle-laden turbulent flows.
{\em Phys. Fuids\/} 28:033302 (2016).

\bibitem[Ni \& Xia(2013)]{Ni2013}
{\sc Ni R. \& Xia K.-Q.} 
Experimental investigation of pair dispersion with small initial separation in convective turbulent flows
{\em Phys. Rev. E\/} {87}, 063006 (2013).

\bibitem[Nicolleau \& Nowakowski(2011)]{Nicolleau2011}
{\sc Nicolleau, F. C. G. A. \& Nowakowski, A. F.} 
Presence of a Richardson's regime in kinematic simulations.
{\em Phys. Rev. E\/} {83}, 056317 (2011).

\bibitem[Obukhov(1941)]{Obukhov1941}
{\sc Obukhov A.} 
Spectral energy distribution in a turbulent flow.
{\em Izv. Akad. Xauk. SSSR. Ser. Geogr. i Geojz\/} 5:453--466. (Translation : Ministry of Supply. p. 21 1097) (1941).

\bibitem[Pedrizzetti \& Novikov(1994)]{Pedrizzetti1994}
{\sc Pedrizzetti G. \& Novikov E. A.} 
On Markov modelling of turbulence. 
{\em J. Fluid Mech.} 280:69–93, (1994).

\bibitem[Richardson(1926)]{Richardson1926}
{\sc Richardson L. F.} Atmospheric diffusion shown on a distance-neighbour graph.
{\em Proc. Roy. Soc. Lond. A\/} 100:709-737 (1926).

\bibitem[Salazar \& Colins(2009)]{Salazar2009}
{\sc Salazar, J. P. L. \& Collins, L. R.}
Two-Particle Dispersion in Isotropic Turbulent Flows.
{\em Annu. Rev.  Fluid Mech.\/} {41}, 405--432, (2009).

\bibitem[Sawford(2001)]{Sawford2001}
{\sc Sawford, B.} 
Turbulent Relative Dispersion.
{\em Annu. Rev.  Fluid Mech.\/} 33:289--317. (2001).

\bibitem[Sawford et al(2008)]{Sawford2008}
{\sc Sawford B. L., Yeung P. K., Hackl J. F.}
Reynolds number dependence of relative dispersion statistics in isotropic turbulence. 
{\em Phys. Fluids} 20:065111, (2008).

\bibitem[Scatamacchia et al(2012)]{Scatam2012}
{\sc R. Scatamacchia, L. Biferale \& F. Toschi}
Extreme Events in the Dispersions of Two Neighboring Particles Under the Influence of Fluid Turbulence
{\em Phys. Rev. Lett.} 109:144501, (2012).

\bibitem[Tampieri(2017)]{Tampieri2017}
{\sc Tampieri F.}
Turbulent Dispersion. In: Turbulence and Dispersion in the Planetary Boundary Layer.
{\em  Physics of Earth and Space Environments. Springer, Cham.} pp: 155-189, (2017)

\bibitem[Tatarski(1960)]{Tatarski1960}
{\sc Tatarski V. I.}
Radiophysical methods for investigating atmospheric turbulence.
{\em Izv.Vyssh. Uchebn. Zaved. Radiofiz}. 3:551–83, (1960).

\bibitem[Taylor(1921)]{Taylor1921}
{\sc Taylor G. I.}
Diffusion by continuous movements.
{\em Proc. Lond. Math. Soc.}. 20, 196–211, (1921).

\bibitem[Thalabard et al(2014)]{Thalabard2014}
{\sc Thalabard S., Krstulovic G. \& Bec J.}
Turbulent pair dispersion as a continuous-time random walk 
{\em J.~Fluid Mech.\/} {755}, R4:1--12 (2014).

\bibitem[Thomson(1996)]{Thomson1996}
{\sc \sc Thomson, D. J.}
The separation of particlepairs in the eddy damped quasinormal Markovian approximation. 
{\em Phys. Fluids} 8:642–44, (1996).

\bibitem[Thomson \& Devenish(2005)]{Thomson2005}
{\sc \sc Thomson, D. J. \& Devenish, B. J.}
Particle pair separation in kinematic simulations. 
{\em J.~Fluid Mech.\/} {526}, 277--302 (2005).

\bibitem[Tsuji(2004)]{Tsuji2004}
{\sc Tsuji Y.} 
Intermittency effect on energy spectrum in high-Reynolds number turbulence. 
{\em Phys. Fluids}, 16: L43 (2004).

\bibitem[Tsuji(2009)]{Tsuji2009}
{\sc Tsuji Y.} 
High-Reynolds-number experiments: the challenge of understanding universality in turbulence. 
{\em Fluid Dyn. Res.}, 41(6):064003 (2009).

\bibitem[Turfus(1987)]{Turfus1987}
{Turfus C. \& Hunt J. C. R. }
A stochastic analysis of the displacements of fluid element in inhomogeneous turbulence using Kraichnan's method of random modes.
{\em In: Comte-Bellot G., Mathieu J. (eds) Advances in Turbulence}, 
Springer, Berlin, Heidelberg, 191-203 (1987). doi: 10.1007/978-3-642-83045-7\_23.

\bibitem[Wolf et al(2004)]{Wolf2004}
{\sc Wolf M. C., Voigt R., \& Moore P. A.}
Spatial arrangement of odor sources modifies the temporal aspects of crayfish search strategies.
{\em J. Chem. Ecology\/} 30(3):501--517 (2004).

\bibitem[Yeung \& Borgas(2004)]{Yeung2004}
{\sc Yeung P. K. \& Borgas M. S.}
Relative dispersion in isotropic turbulence. Part 1. Direct numerical simulations and Reynolds-number dependence. 
{\em J. Fluid Mech.} 503:93–124, (2004).

\bibitem[Virant \& Dracos(1997)]{Virant1997}
{\sc Virant M. \& Dracos, T.}
3D PTV and its application on Lagrangian motion. 
{\em Meas. Sci. Technol.} 8:(12), 1539–1552.

}
\end{thebibliography}
\end{document}